\documentclass[11pt]{article}

\usepackage[utf8]{inputenc}
\usepackage[T1]{fontenc}
\usepackage{libertine} 
\usepackage{authblk}

\usepackage{amsfonts,amsmath,amssymb,amsthm}
\usepackage{mathtools}
\usepackage{nicefrac,xfrac}
\usepackage{bm}

\usepackage[margin=1in]{geometry}
\usepackage{changepage}
\usepackage{float}
\usepackage{array}
\usepackage{booktabs}
\usepackage{caption}
\usepackage{subcaption}
\usepackage{multirow}

\usepackage[dvipsnames]{xcolor}
\definecolor{HarvestOrange}{HTML}{C9653C}
\definecolor{HarvestBrown}{HTML}{786662}
\definecolor{HarvestGreen}{HTML}{475A50}
\definecolor{ChocolateBrown}{HTML}{765457}
\definecolor{AppleLeafGreen}{HTML}{5B8258}

\usepackage{graphicx}

\usepackage{tikz}
\usepackage{pgfplots}
\pgfplotsset{compat=1.18}
\usetikzlibrary{
    patterns,intersections,arrows.meta,positioning,
    arrows,decorations.markings
}
\usepgfplotslibrary{fillbetween}

\tikzstyle{vecArrow} = [
    thick,
    decoration={markings,mark=at position 1 with {\arrow[semithick]{open triangle 60}}},
    double distance=1.4pt, shorten >=5.5pt,
    preaction={decorate},
    postaction={draw,line width=1.4pt,white,shorten >=4.5pt}
]
\tikzstyle{innerWhite} = [
    semithick, white,line width=1.4pt,shorten >=4.5pt
]

\usepackage{enumitem}
\usepackage{accents}
\usepackage{comment}
\usepackage{csquotes}

\usepackage{natbib}

\usepackage[
    colorlinks=true,
    linkcolor=blue!70!black,
    citecolor=blue!70!black,
    urlcolor=black,
    breaklinks=true
]{hyperref}
\usepackage{cleveref} 

\usepackage{listings}
\usepackage[ruled,vlined,linesnumbered]{algorithm2e} 

\SetAlFnt{\small}
\SetAlCapFnt{\small}
\SetAlCapNameFnt{\small}
\SetAlCapHSkip{0pt}
\IncMargin{-\parindent}

\usepackage{tcolorbox}

\usepackage{thmtools,thm-restate}

\theoremstyle{plain}

\theoremstyle{definition}
\newtheorem{definition}{Definition}[section]
\newtheorem{example}[definition]{Example}



\newcommand{\AutoAdjust}[3]{\mathchoice{ \left #1 #2  \right #3}{#1 #2 #3}{#1 #2 #3}{#1 #2 #3} }
\newcommand{\inParentheses}[1]{\AutoAdjust{(}{#1}{)}}
\newcommand{\inBrackets}[1]{\AutoAdjust{[}{#1}{]}}

\DeclareMathOperator{\argmax}{argmax}

\newcommand{\subjectTo}{\mathrm{s.t.}\quad}

\newcommand{\condition}{\mid}

\newcommand{\prob}[2][]{\mathrm{Pr}\ifthenelse{\not\equal{}{#1}}{_{#1}}{}\!\left[{#2}\right]}
\newcommand{\expect}[2][]{\mathbb{E}\ifthenelse{\not\equal{}{#1}}{_{#1}}{}\!\left[{#2}\right]}
\newcommand{\norm}[2]{\left\|{#2}\right\|_{#1}}
\newcommand{\lnorm}[1]{\mathcal L_{#1}}

\newcommand\bbR{\mathbb R}

\newcommand{\reals}{\bbR}
\newcommand{\probSpace}{\Delta}

\newcommand{\abs}[1]{\left| #1 \right|}

\newcommand{\query}{q}
\newcommand{\orgResponse}[1][]{z_{#1}}
\newcommand{\newResponse}[1][]{\tilde{z}_{#1}}
\newcommand{\spaceIns}[1][]{y_{#1}}

\newcommand{\numSlot}{M}
\newcommand{\slot}{j}

\newcommand{\numIns}{K}

\newcommand{\numAd}{N}
\newcommand{\ad}{i}
\newcommand{\adCre}[1][\ad]{a_{#1}}

\newcommand{\numGen}{L}
\newcommand{\genre}{g}
\newcommand{\genreSpace}{\mathcal{G}}

\newcommand{\intent}{t}
\newcommand{\intentSpace}{\mathcal{T}}

\newcommand{\newagentvar}[2]{%
\expandafter\newcommand\expandafter{\csname #1\endcsname}[1][\ad]{{#2}_{##1}}%
\expandafter\newcommand\expandafter{\csname #1s\endcsname}[1][]{\boldsymbol{#2}_{##1}}%
\expandafter\newcommand\expandafter{\csname #1Es\endcsname}[1][\ad]{\tilde{#2}_{##1}}%
\expandafter\newcommand\expandafter{\csname #1sEs\endcsname}[1][]{\tilde{\boldsymbol{#2}}_{##1}}%
\expandafter\newcommand\expandafter{\csname #1sOPT\endcsname}[1][]{\boldsymbol{#2}^*_{##1}}%
\expandafter\newcommand\expandafter{\csname #1OPT\endcsname}[1][]{#2^*_{##1}}%
}

\newagentvar{val}{v}
\newagentvar{wel}{w}
\newagentvar{bid}{b}
\newagentvar{alloc}{x}
\newagentvar{pay}{p}
\newagentvar{util}{u}
\newagentvar{coh}{c}
\newagentvar{sco}{s}
\newcommand{\totalWel}{W}
\newcommand{\totalWelEs}{\tilde{W}}

\newcommand{\valMax}{\bar{v}}

\newcommand{\gapVal}{\varepsilon_{\mathrm{V}}}
\newcommand{\gapCoh}{\varepsilon_{\mathrm{C}}}
\newcommand{\gapDis}{\varepsilon_{d}}

\newcommand{\cohsEsMat}{\tilde{\bm C}}
\newcommand{\welsMat}{\bm W}

\newcommand{\valsEsMat}{\tilde{\bm V}}

\newcommand{\TVD}{d_{\mathrm{TV}}}
\newcommand{\Dist}[1][]{D_{#1}}

\newcommand{\cali}{f}

\newcommand{\bidV}{X}

\title{Ad Insertion in LLM-Generated Responses}

\author[1]{Shengwei Xu\thanks{Shengwei Xu and Zhaohua Chen contribute equally to this work.}\thanks{Shengwei Xu and Grant Schoenebeck are supported by NSF Grant No.~2313137.}}
\author[2]{Zhaohua Chen\protect\footnotemark[1]}
\author[2]{Xiaotie Deng}
\author[3]{Zhiyi Huang\thanks{Zhiyi Huang and Grant Schoenebeck are corresponding authors.}}
\author[1]{Grant Schoenebeck\protect\footnotemark[2]\protect\footnotemark[3]}

\affil[1]{University of Michigan\\ \textit{\{shengwei}, \textit{schoeneb\}@umich.edu}}
\affil[2]{Taobao \& Tmall Group of Alibaba, Renmin University of China\\ \textit{chenzhaohua.czh@alibaba-inc.com}}
\affil[3]{City University of Hong Kong\\ \textit{csdeng@cityu.edu.hk}}
\affil[4]{The University of Hong Kong\\ \textit{zhiyi@cs.hku.hk}}

\date{}

\begin{document}

\maketitle

\begin{abstract}
Sustainable monetization of large language models (LLMs) remains a critical open challenge. Traditional search advertising, which relies on static keywords, fails to capture the fleeting, context-dependent user intent---the specific information, goods, or services a user seeks---embedded in conversational flows. Beyond the standard goal of \emph{social welfare maximization}, effective LLM advertising requires \emph{contextual coherence} (aligning ads semantically with transient user intent), \emph{computational efficiency} (avoiding user-facing latency), and adherence to \emph{ethical and regulatory standards}, including privacy preservation and explicit ad disclosure. Although recent solutions have explored bidding at the token and query levels, neither category holistically satisfies these constraints.

We propose a framework that resolves these tensions through two decoupling strategies. First, we \emph{decouple ad insertion from response generation} to facilitate pre-screening and explicit disclosure. Second, we \emph{decouple bidding from specific user queries by using ``genres''} (high-level semantic clusters) as a proxy. This allows advertisers to bid on stable categories rather than sensitive real-time responses, reducing computational burden and privacy risks. Applying the VCG auction mechanism to this genre-based framework provides approximate guarantees for dominant-strategy incentive compatibility (DSIC), individual rationality (IR), and social welfare. In synthetic experiments with $10^5$ advertisers and 100 candidate slots, VCG clears in approximately 1.25 seconds on a consumer-grade laptop. Finally, we introduce an ``LLM-as-a-Judge'' metric for estimating contextual coherence. Its predictions correlate with mean human ratings at Spearman's $\rho\approx 0.66$ and have a higher correlation with the leave-one-out group mean than 29 of 36 individual raters (80.6\%).
\end{abstract}

\section{Introduction}
\label{sec:intro}

ChatGPT now serves more than 900 million weekly active users and has over 50 million consumer subscribers~\citep{openai2026scaling}.
Given the considerable cost of training and deploying LLMs---a recent International Data Corporation (IDC) forecast projected \$108 billion in U.S.\ GenAI spending in 2028~\citep{idc2024worldwide}---sustainably monetizing services on which so many users rely remains challenging.

A natural approach to monetizing LLM services is to integrate ads into LLM responses, similar to current search engines. 
Indeed, a recent survey~\citep{rainie2025close} revealed that 68\% of users have used LLMs for ``searching to find facts quickly, like a search engine'', and 57\% of them have used LLMs for ``getting information about products and services''.
For context, Google's advertising revenue exceeded \$260 billion in 2024~\citep{alphabet2024annualreport}.
Thus, introducing ads into LLM responses could help defray the costs of these services and is particularly relevant to conversational commerce.

However, the transition from ``Search'' to ``Chat'' is not merely a shift in interface; it presents unique structural challenges.
In a search engine, user intent \citep{broder2002taxonomy} is naturally encoded by static keywords, so advertisers can bid on keywords to target a particular intent. These keywords indicate users' state of mind and, in particular, the goods or services they may be seeking. In an LLM chat, the analogous user intent may shift throughout a complex conversational flow.
For example, an LLM response about a user's family vacation to New York City might start with a discussion of attractions, then move on to accommodations, transportation, and budget considerations.
Ideally, an ad about transportation would appear surrounded by text about transportation rather than text regarding accommodations.
Placing an ad in the right location not only improves its effectiveness but also makes it less disruptive to the user experience.

This fundamental shift raises three technical challenges:
\begin{enumerate}
    \item \emph{Contextual coherence}: How do we identify the evolving user intent in the LLM response to provide coherent ads, so that we can align advertisers' welfare with specific users' intents?
    \item \emph{Computational efficiency}: How do we insert ads without introducing latency that degrades the real-time user experience?
    \item \emph{Social welfare}: How can we design a mechanism that approximately maximizes social welfare, defined as the sum of advertiser and platform utilities, in the absence of explicit, static keywords on which to bid?
\end{enumerate}

In addition, inserting commercial bias into the organic flow is constrained by \emph{practical considerations}. 
For example, on the \emph{regulation} side, LLM platforms must explicitly mark where ads are inserted, to align with the U.S.\ Federal Trade Commission (FTC) policy that ``advertising and promotional messages that
are not identifiable as advertising to consumers are deceptive if they mislead consumers into
believing they are independent, impartial, or not from the sponsoring advertiser itself'' \citep{ftc2015enforcement}. 
On the \emph{ethics} side, ads in the LLM response should also be non-deceptive and non-misleading.
Further, if advertisers bid at the level of each query or token, they may gain access to potentially sensitive user prompts, creating \emph{privacy} risks.

With these challenges in mind, we aim to answer the following research question:

\begin{center}
    \textit{How can we design a practical framework for integrating ads into LLM responses while preserving advertiser welfare, contextual coherence, and computational efficiency?}
\end{center}

\subsection{Existing Solutions and Challenges}\label{sec:tech-chal}

Traditional search and display advertising models, which rely on fixed ad slots and static user intent, do not generalize well to this domain. LLM responses are dynamic and lack predetermined slots, and an ad's effectiveness depends heavily on its \emph{contextual coherence} with the surrounding text and the user's evolving intent. Conceptually, this resembles \emph{native advertising}, in which ads are woven into a narrative. However, native ads are typically prepared offline and are labor-intensive. Because LLM platforms operate in real time, an automated mechanism must achieve contextual integration with high \emph{computational efficiency}.

In the past two years, there has been a surge of studies on integrating ads into LLM responses. A position paper by \citet{feizi2023online} identifies several critical components of LLM advertising, including the question of ``what to bid for.''
To this end, existing works can be roughly classified into two groups, both of which face challenges regarding computational efficiency and practical concerns that our framework addresses.

The first group~\citep{duetting2024mechanism,hajiaghayi2024ad,zhao2025llm,han2026mechanism} allows advertisers to bid at the \emph{token or segment level}. Here, advertisers bid repeatedly during generation for their content to appear as the next token. This approach introduces significant latency. 
More critically, as reported by \citet{zelch2024user}, if LLMs are prompted to blend ads subtly into the organic response, the resulting ads can be difficult to recognize.
Because the content is generated in real time by an LLM, it is also susceptible to hallucinations and misinformation.

The second group~\citep{soumalias2024truthful,dubey2024auctions,mordo2024sponsored,balseiro2025position} focuses on bidding at the \emph{query or response level}. While it avoids token-level generation issues by inserting predefined ads, bidding on the entire query forces advertisers to value the interaction in aggregate, preventing them from targeting specific moments within a response (e.g., a travel plan containing both ``flight'' and ``hotel'' contexts).
Furthermore, this approach typically requires advertisers to submit a bid for each specific user query. This also imposes a massive computational burden on advertisers, who must perform real-time semantic analysis to value each unique query effectively. 

Crucially, both groups of approaches typically expose sensitive user prompts to advertisers, raising significant privacy concerns.

\subsection{Exploratory Survey Results} \label{sec:survey-results}

To explore users' concerns regarding LLM ads, we conducted an exploratory survey with 48 participants in April 2025.\footnote{Participant demographics are provided in \Cref{app:exp-setup}.}

A majority (56.3\%) expected ads to appear in LLM responses within a year, whereas only 12.5\% thought this would not happen. Despite this expectation, the acceptability of such ads remained low: only 4.2\% rated them as acceptable, whereas 64.6\% rated them as unacceptable.




To understand this resistance, we analyzed open-ended responses regarding participants' concerns.\footnote{Non-English responses were translated into English using Claude 4.5 Sonnet and verified by the authors.}
Three major themes emerged.
First, over 37.5\% of responses specifically mentioned \emph{trust and objectivity} issues.
Users worried about unconscious manipulation through ``hidden advertisements'':

\begin{quote}
\em
``For some hidden advertisements, due to our trust in AI, we often fail to notice their existence, leading to unconscious biased choices. This is very scary.''
%
\end{quote}

Second, 20.8\% of participants expressed concern that using LLMs to generate promotional content may lead to \emph{misinformation or hallucinations} about the advertised product, raising questions about liability and oversight that rarely arise in traditional advertising.

\begin{quote}
\em
``We need to pay more attention to the authenticity of AI-generated ads, and clarify any potential false advertising and who's responsible for it.''
%
\end{quote}

Third, 12.5\% emphasized the importance of \emph{contextual relevance} and a seamless user experience.

\begin{quote}
\em
``Will unrelated ads appear in AI response?''
\end{quote}

Together, the survey evidence indicates that users expect LLM ads soon but will only condone them when \emph{clearly disclosed}, \emph{responsible}, and \emph{contextually coherent}.
This aligns with legal guidance, reinforces the need to explicitly mark ads in LLM responses, and further motivates our framework.

\subsection{Our Contributions}

Our primary contribution is a practical framework for integrating ads into LLM responses with high economic welfare, high contextual coherence, and high computational efficiency. 
En route to this goal, we introduce two layers of decoupling, respectively answering the question of ``how ads are displayed'' and ``what to bid for'' raised by \citet{feizi2023online}.

\begin{figure}[ht]
    \centering
    \includegraphics[width=1\linewidth]{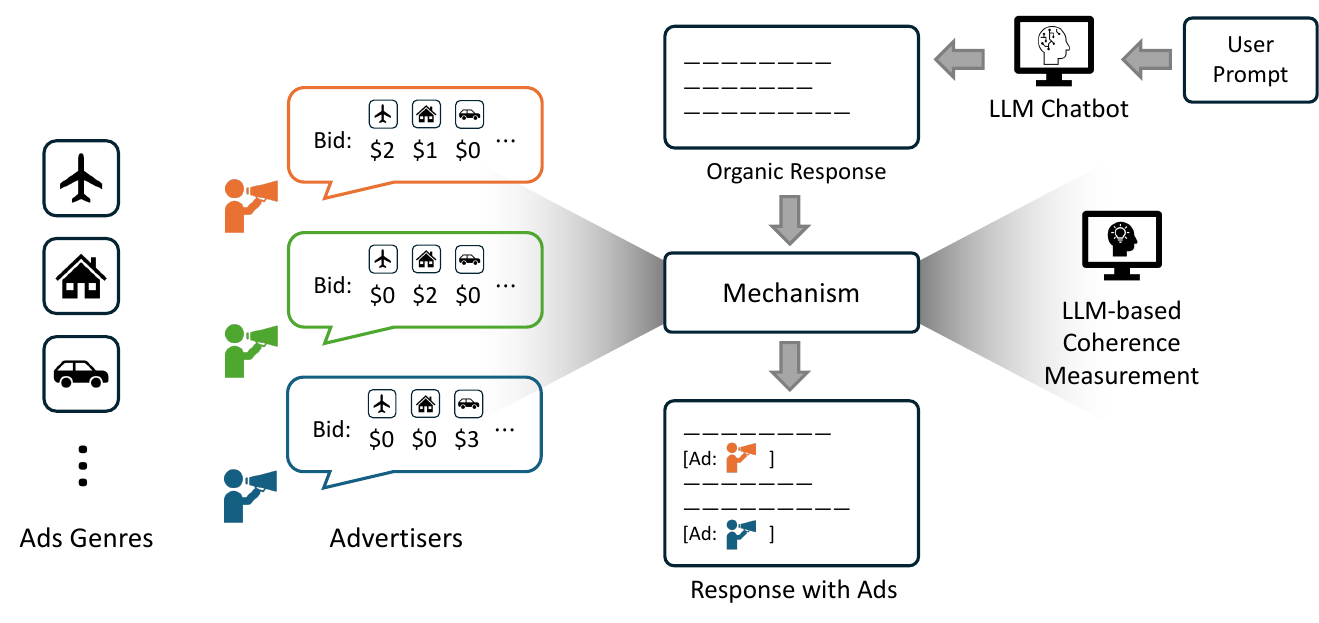}
    \caption{Our framework for ad insertion in LLM-generated responses.}
    \label{fig:framework}
\end{figure}

Concerning how ads are displayed, we \textbf{decouple ad insertion from response generation}.
In our framework (see \Cref{fig:framework}), upon receiving a query, an LLM chatbot first generates an organic response to the query, and independently, advertisers compete for the right to insert their pre-submitted ads in different positions in the organic response without further modification (e.g., at the end of a paragraph).
Such a decoupling comes with two practical advantages:
(1) \emph{Regulation and ethics} concerns related to LLM ads are addressed at their root. 
Since ad creatives typically stay unchanged for weeks, the service provider can pre-screen and block deceptive content before integrating them into LLM responses.
Any ad appearance can also be explicitly marked in the response.
(2) The decoupled response-generation and mechanism-design modules can be \emph{independently optimized} with minimal coordination cost. 
For example, classic auction mechanisms, such as VCG, can be seamlessly migrated into the new scenario. 

\begin{figure}[ht]
    \centering
    \includegraphics[width=0.8\linewidth]{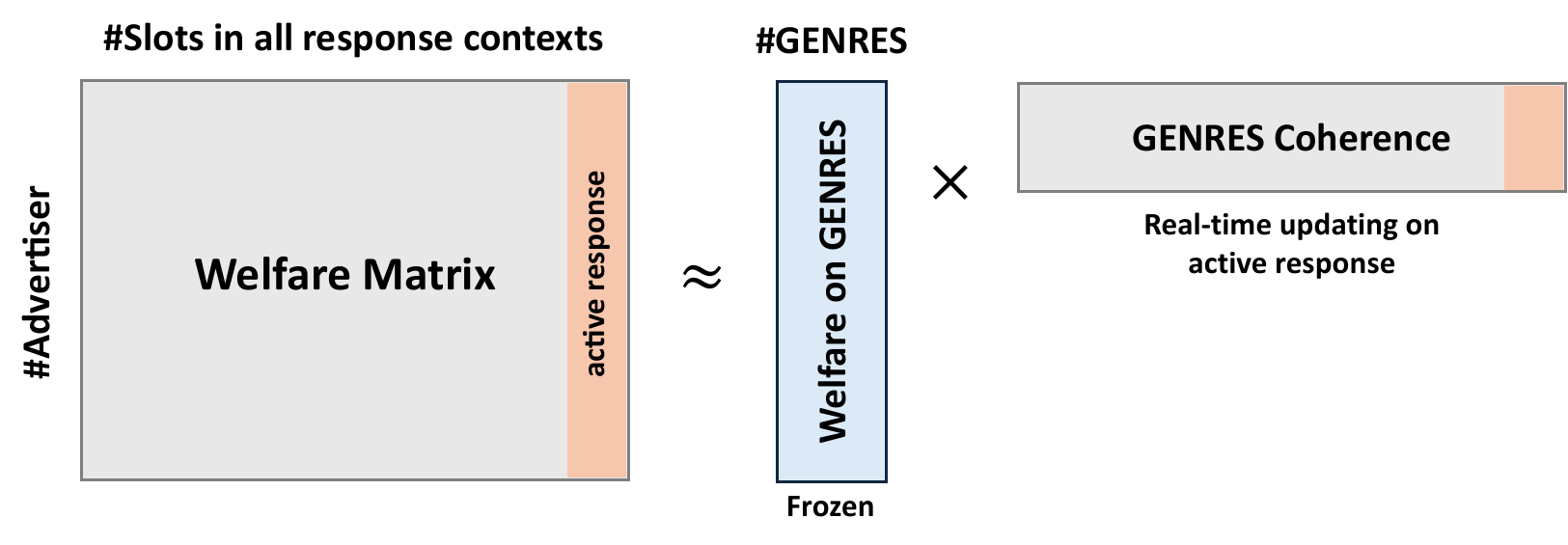}
    \caption{An illustration of how genres work under the matrix decomposition perspective.}
    \label{fig:welfare_matrix_fac}
\end{figure}

The other layer of decoupling operates on the bidding interface, where we \textbf{decouple bidding from the response context}.
Recall that, compared to search ads that rely on keywords, in LLM responses, user intent evolves and thus will be slot-dependent. As we discussed in \Cref{sec:tech-chal}, it is infeasible for advertisers to bid on every possible slot in the response context. 
Therefore, we introduce the concept of \textit{genres}, high-level semantic clusters (e.g., \emph{hotels}, \emph{airlines}, \emph{food}), as a tractable proxy for the underlying user \textit{intents}.
Advertisers bid offline on stable, predefined genres.
At inference time, the platform only estimates a \emph{coherence} probability between each genre and each candidate slot in the conversation.
Combining (1) an advertiser's genre bids and (2) slot-level genre coherence, we estimate the welfare from assigning that advertiser to a slot by weighting the bids by their coherence probabilities. Conceptually, this is analogous to decomposing a welfare matrix into advertiser--genre and genre--slot factors (see \Cref{fig:welfare_matrix_fac}).

Such a design decouples the advertisers from the user's response context, yielding three immediate advantages:
(1) \emph{Efficiency:} advertisers no longer need to value every query or token in real time, and the platform runs auctions over a tractable predefined genre set rather than a massive space of queries or tokens;
(2) \emph{Privacy:} advertisers bid on coarse categories rather than on potentially sensitive user prompts/responses;
(3) Practically, because genres play a role analogous to keywords, the approach can be integrated with existing search-ad infrastructure with limited modification.

However, an estimation error arises because genres may not capture all information about user intent and therefore may not perfectly recover true valuations.
This error shrinks as genres become more fine-grained clusters and as coherence estimation becomes more accurate. We now characterize how this approximation affects mechanism design guarantees.

\paragraph{Mechanism design.}
We analyze the economic properties of this genre-based auction. A core challenge is that genres are a coarse proxy of true user intent; therefore, the platform cannot know the advertiser's true valuation for a specific context perfectly. 
Despite this information asymmetry, in \Cref{sec:mechanisms}, we prove that applying the VCG mechanism to the estimated values yields strong theoretical guarantees. Specifically, we show that the mechanism is \emph{approximately} dominant strategy incentive compatible (DSIC) and individually rational (IR). Furthermore, when assuming that advertisers bid truthfully, the mechanism achieves \emph{approximately optimal social welfare}.
We derive error bounds showing that the incentive to deviate from truthful bidding is bounded by the granularity of the genres and the accuracy of the coherence estimation.
In \Cref{sec:prototypes}, we empirically demonstrate that computing the VCG allocation and payment is computationally efficient.

\paragraph{Coherence measurement.}
For coherence estimation, in \Cref{sec:coherence}, we propose two methods based on recent natural language generation (NLG) evaluation techniques: sentence embeddings \citep{devlin2019bert,zhang2019bertscore} and LLM-as-a-Judge \citep{bai2024benchmarking,zheng2023judging}.

To evaluate their reliability, in \Cref{sec:experiment-coherence}, we conduct empirical studies comparing these coherence signals with human-annotated ratings. 
Our results show that DeepSeek-R1 and GPT-5-based judges achieve Spearman correlations of approximately 0.66 with the mean ratings of 36 participants.
The GPT-5 judge has a higher correlation with the leave-one-out group mean than 29 of 36 individual raters (80.6\%).
We further observe that the correlation of LLM-as-a-Judge metrics improves consistently with model capability, suggesting continued gains as foundation models advance. We provide all code, experiment materials, and anonymized survey data in the supplementary material.

\section{Framework}

In this section, we introduce our framework for ad insertion in LLM-generated responses. As shown in \Cref{fig:framework}, when a user sends a query $\query$ to the LLM service, the LLM first generates an ad-free organic response $\orgResponse$. Independently, advertisers submit their ad creatives and bidding information to the LLM service provider. Based on the advertisers' values, the mechanism outputs a final response $\newResponse$ that preserves the sentences and their order in $\orgResponse$ while inserting several predefined ad creatives among them.

Crucially, our framework decouples the insertion of ads from the generation of the response. This separation circumvents the risk of hallucinated ad content and ensures transparency, preventing the subtle blending of ads with organic responses.

Our framework consists of three core modules: 
\begin{enumerate}
\item A valuation model in which advertisers estimate utility using ``genres'' as a proxy for latent user intents. This design isolates private user queries from advertisers and relieves advertisers of the burden of bidding on every individual query.
\item A contextual-coherence estimator that measures the alignment between a given genre and the organic response.
\item An auction mechanism that combines genre-based bids with the platform's contextual-coherence estimates to determine ad allocations and payments.
\end{enumerate}

\paragraph{Organic response and ad slots.}
Formally, we model the organic response as a sequence of $\numSlot+1$ text segments, $\orgResponse = \orgResponse[1]\orgResponse[2]\cdots\orgResponse[\numSlot+1]$. For $\slot \in \{1,\ldots,\numSlot\}$, let $\spaceIns[\slot]$ denote the candidate ad slot between $\orgResponse[\slot]$ and $\orgResponse[\slot+1]$.
Throughout our framework, we consider a single organic response $\orgResponse$ to a specific user query. We therefore use $\slot$ as shorthand for the semantic context at that position in the current response.

\subsection{Advertisers' Valuation and Genres}
\label{subsec:valuation}

We now model the advertisers' valuation structure. Consider $\numAd$ advertisers, where each advertiser $\ad \in \{1,\ldots,\numAd\}$ holds an ad creative $\adCre$. Let $\wel[\ad, \slot]$ be a random variable representing the value derived by the advertiser from an impression at slot $\slot$.

\begin{figure}[!ht]
    \centering
    \includegraphics[width=1\linewidth]{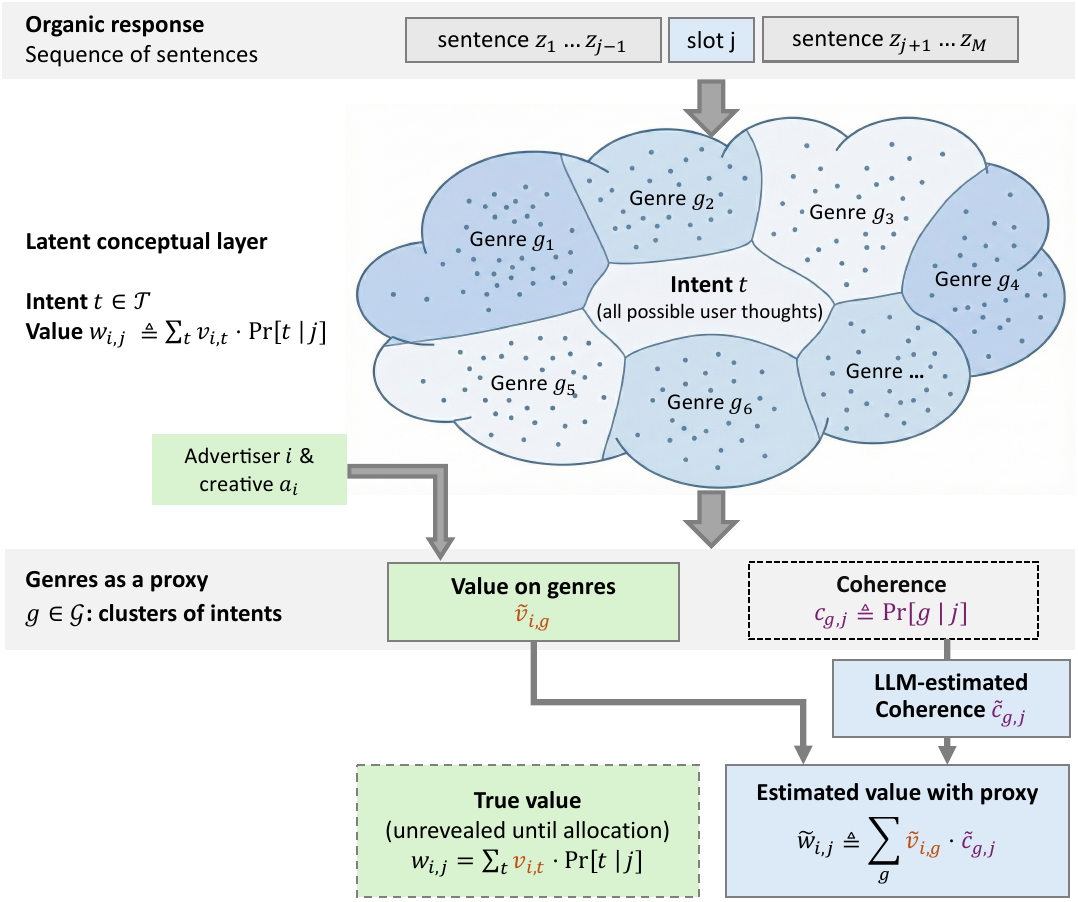}
    \caption{Our model for advertisers' valuation and platform's coherence with genres.\\
    \small
    A genre acts as a bucket for multiple intents. The platform uses the average value $\valEs[\ad, \genre]$ as a static proxy, since the true value $\val[\ad, \intent]$ is impossible to process due to the extremely large intent space $\intentSpace$. \\
    Relying on static proxy averages over multiple intents introduces an estimation error (see \Cref{eg:genre_proxy_error}). This error decreases as the granularity of the genre set $\genreSpace$ increases; in the extreme case where genres equal intents, the proxy matches the true value.}
    \label{fig:genre_model}
\end{figure}

\paragraph{Intents and advertiser valuation.} 

The value $\wel[\ad, \slot]$ depends on the specific context and the user's intent at the slot. We capture this dependency using the concept of an \emph{intent}, denoted by $\intent \in \intentSpace$, where $\intentSpace$ is the set of all possible intents. An intent precisely encapsulates the user's latent thought process while reading the response at a specific slot.
However, the textual context at slot $\slot$ does not fully reveal the user's private latent factors (e.g., budget constraints, current mood). To capture the randomness arising from this lack of full knowledge, we use $\prob{\intent \condition \slot}$ to represent the probability that intent $\intent$ is active given the context observed at slot $\slot$.

\begin{example}
Consider a user named Alice querying an LLM for a travel itinerary to New York City. Upon reading the transportation section of the response ($\slot$), Alice's true intent depends on latent factors unknown to the platform. She might prefer a ``first-class flight'' if she recently received a bonus or a ``Greyhound coach'' if she is budget-constrained. The distribution $\prob{\intent \condition \slot}$ captures this uncertainty.
\end{example}

Conditioning on the intent $\intent$, the value of an ad creative $\adCre$ is independent of the slot index $\slot$.
Let $\val[\ad, \intent] \in [0, \valMax]$ represent advertiser $\ad$'s value given user intent $\intent$. Here, $\valMax$ is an upper bound on an advertiser's value. Thus, advertiser $\ad$'s \emph{ground-truth} value at slot $\slot$ is
\begin{align}
    \wel[\ad, \slot] &\triangleq \sum_{\intent} \val[\ad, \intent]\cdot \prob{\intent \condition \slot}. \label{eq:welfare formula with intents}
\end{align}

However, the challenge lies in the cardinality of $\intentSpace$, since intents cover the infinite nuances of natural language conversations. Directly estimating $\prob{\intent \condition \slot}$ or bidding on individual $\val[\ad, \intent]$ is computationally intractable.

\paragraph{Genres as a proxy.}
To resolve this intractability, we introduce \emph{genres} as a tractable proxy. We model a genre $\genre \in \genreSpace$ as a set of intents, where $\genreSpace$ is a predefined finite set of $\numGen$ genres that forms a partition of the intent space $\intentSpace$. Thus, genres do not intersect, and their union covers all intents. This formulation aligns with industry standards for contextual targeting, such as the IAB Content Taxonomy \citep{iab_taxonomy}, which groups similar products into verticals (e.g., food, automotive, and insurance).
The probability that genre $\genre$ aligns with the context at slot $\slot$ is $\prob{\genre \condition \slot} = \sum_{\intent \in \genre} \prob{\intent \condition \slot}$.

We assume that the advertiser knows $\valEs[\ad, \genre]\in [0, \valMax]$, the average realized value of its ad creative across a reference distribution of contexts in which genre $\genre$ has positive probability.
In practice, this assumption is supported by the fact that advertisers are required to bid on keywords for search engine ads. Thus, the advertiser's value in \Cref{eq:welfare formula with intents} can be approximated as
\begin{align}
    \wel[\ad, \slot] \approx \sum_{\genre} \valEs[\ad, \genre]\cdot \prob{\genre \condition \slot}. \label{eq:welfare formula with genres}
\end{align}

\begin{example}\label{eg:genre_proxy_error}
Consider ``Airlines'' as a broad genre acting as a proxy for specific user intents, such as ``Business Class'' and ``Economy''.
While the broad genre remains constant, the underlying intent varies by context: business travel implies a high probability of ``Business Class,'' whereas a budget family vacation implies ``Economy.''
This aggregation causes a valuation error because advertisers value these intents differently. For instance, a low-cost carrier derives high utility from ``Economy'' but low utility from ``Business Class''.
Since the bid is tied to the coarse genre, it remains static regardless of the slot's specific content. Consequently, the valuation fails to leverage the business or budget signal present in the context, leading to a mismatch between the proxy and true values.
\end{example}

As the granularity of genres increases, the gap between $\sum_{\intent} \val[\ad, \intent] \cdot \prob{\intent \condition \slot}$ and $\sum_{\genre} \valEs[\ad, \genre]\cdot \prob{\genre \condition \slot}$ decreases, since each $\genre$ captures more information about the context of $\slot$. 
In the extreme case where the number of genres equals the number of intents, each genre represents one intent and the approximation error becomes zero.
In \Cref{app:low-rank}, we provide intuition connecting our genre-based model to a low-rank approximation of the welfare matrix.
In \Cref{sec:mechanisms}, we formally discuss how the approximation error impacts the mechanism's truthfulness and social welfare maximization.

\subsection{Contextual Coherence}
\label{sec:coherence_model}

A crucial component of the welfare equation \eqref{eq:welfare formula with genres} is the term $\prob{\genre \condition \slot}$ for all genres $\genre$. 
It captures the probability that genre $\genre$ aligns with the context of the organic response at position $\slot$.
Intuitively, if a genre is highly correlated with the text surrounding a slot, this probability increases. We call this probability \emph{coherence} and denote it by $\coh[\genre, \slot] \triangleq \prob{\genre \condition \slot}$.
We further denote $\cohs[\slot] \triangleq (\coh[\genre, \slot])_{\genre \in \{1, 2, \cdots, \numGen\}}$. 

Contextual coherence has been studied in online advertising for decades under names such as \emph{semantic matching} and \emph{relevance modeling} \citep{ribeiro2005impedance,grbovic2016scalable}. Its widespread industrial use suggests that $\coh[\genre, \slot]$ can be estimated efficiently by building on prior methods.

In our framework, coherence is measured on the platform side since only the platform holds the organic response. 
We use $\cohEs[\genre, \slot]$ to denote the platform's estimation of $\coh[\genre, \slot]$, and analyze how the estimation error impacts the theoretical bounds in \Cref{sec:mechanisms}. 
In \Cref{sec:coherence}, we propose two prototype methods for estimating coherence with modern language models.

\subsection{Auction Mechanism}

Finally, we formally define the auction mechanism. As illustrated in \Cref{fig:genre_model}, the platform conducts the auction using the advertisers' genre-based bids and its own coherence estimates.

\paragraph{Bids and Allocation.}
Each advertiser $\ad$ submits a bid vector $\bids[\ad] \triangleq (\bid[\ad, \genre])^\top_{\genre \in \genreSpace}$, where $\bid[\ad, \genre] \in [0, \valMax]$ represents the advertiser's estimated value for an impression within genre $\genre$.
We say an advertiser \emph{truthfully} bids if $\bids[\ad] = \valsEs[\ad] \triangleq (\valEs[\ad, \genre])^\top_{\genre \in \{1, 2, \cdots, \numGen\}}$. 
The platform determines an allocation $\alloc[\ad, \slot] \in \{0, 1\}$, indicating whether ad $\adCre$ is placed in slot $\slot$, and calculates a payment $\pay[\ad] \geq 0$. Let $\numIns \leq \min\{\numAd,\numSlot\}$ be a predetermined integer denoting the number of ads to insert. The allocation must satisfy the following feasibility constraints:
\begin{enumerate}
    \item Each slot contains at most one ad: $\sum_{\ad} \alloc[\ad, \slot] \leq 1, \forall \slot$.
    \item Each advertiser receives at most one slot per response: $\sum_{\slot} \alloc[\ad, \slot] \leq 1, \forall \ad$.
    \item Exactly $\numIns$ ads are inserted: $\sum_{\ad, \slot} \alloc[\ad, \slot] = \numIns$.
\end{enumerate}

\paragraph{Ground truth utility.}
Given the allocation and payment, each advertiser $\ad$ receives a \emph{ground-truth} utility $\util$, which is a function of all advertisers' bids $\bids$ because the allocation $\alloc[\ad, \slot]$ and payment $\pay$ depend on $\bids$.
\[\util(\bids) \triangleq \sum_{\slot} \alloc[\ad, \slot]\cdot \wel[\ad, \slot] - \pay.\]

\paragraph{Proxy utility.}
Because the true values are unknown to the platform, the advertiser's ground-truth value may differ from the estimated value. With estimated coherence $\cohsEs[\slot] \triangleq (\cohEs[\genre, \slot])^\top_{\genre \in \{1,\ldots,\numGen\}}$ and genre values $\valsEs[\ad] \triangleq (\valEs[\ad, \genre])^\top_{\genre \in \{1,\ldots,\numGen\}}$, the \emph{proxy} value that advertiser $\ad$ obtains in slot $\slot$ is
\[\welEs[\ad, \slot] \triangleq \sum_{\genre} \valEs[\ad, \genre]\cdot \cohEs[\genre, \slot] = \valsEs[\ad]^\top \cohsEs[\slot], \]
and the \emph{proxy} utility is therefore 
\[\utilEs(\bids) \triangleq \sum_{\slot} \alloc[\ad, \slot] \welEs[\ad, \slot] - \pay = \sum_{\slot} \alloc[\ad, \slot] \valsEs[\ad]^\top \cohsEs[\slot] - \pay. \]

\subsubsection{Estimation Errors}
The gap between the proxy utility $\utilEs$ and the ground truth utility $\util$ stems from two sources: the coarseness of the genre proxy and the platform's imperfect coherence measurement. We formally define these errors below; they are used in \Cref{sec:mechanisms} to analyze the approximation bound for truthfulness and welfare maximization.

\paragraph{Valuation error.}
Recall that the ground truth value is a sum over intents. We can rewrite this value by grouping intents into their respective genres. Let $\val[\ad, \genre, \slot]$ denote the advertiser's \emph{realized value} for genre $\genre$ in the specific context of slot $\slot$:
For $\prob{\genre \condition \slot}>0$, define
\[ \val[\ad, \genre, \slot] \triangleq \expect{\val[\ad, \intent] \mid \intent \in \genre, \slot} = \sum_{\intent \in \genre} \val[\ad, \intent] \cdot \frac{\prob{\intent \condition \slot}}{\prob{\genre \condition \slot}}. \]
When $\prob{\genre \condition \slot}=0$, we set $\val[\ad, \genre, \slot]=\valEs[\ad,\genre]$ by convention; this value has no effect on the ground-truth welfare.
This allows us to express the ground truth value in terms of genres:
\begin{align*}
    \wel[\ad, \slot] & = \sum_{\genre} \val[\ad, \genre, \slot]\cdot \prob{\genre \condition \slot}.
\end{align*}
The \emph{valuation error} arises because the advertiser bids a static value $\bid[\ad, \genre]$ for the genre (which ideally equals $\valEs[\ad, \genre]$), but the realized value $\val[\ad, \genre, \slot]$ varies by slot (e.g., a ``business trip'' slot versus a ``family vacation'' slot within the ``Airlines'' genre).

\begin{definition}[Valuation Error]
    We define $\gapVal$ as the maximum normalized deviation between the static genre value and the context-specific realized value:
    \[\gapVal \triangleq \frac{\max_{\ad, \genre, \slot} \abs{\valEs[\ad, \genre] - \val[\ad, \genre, \slot]}}{\valMax}.\]
    Here, $\valMax$ is an upper bound on the values $\valEs[\ad, \genre]$.
\end{definition}

\paragraph{Coherence error.}
The second source of error is the platform's ability to estimate the coherence of a genre with the slot context.

\begin{definition}[Coherence Error]
    We define $\gapCoh$ as the maximum $\lnorm{1}$ deviation between the estimated coherence vector and the true probability vector over genres:
    \[\gapCoh \triangleq \max_{\slot} \norm{1}{\cohsEs[\slot] - \cohs[\slot]} = \max_{\slot} \sum_{\genre} \abs{\cohEs[\genre, \slot] - \coh[\genre, \slot]}.\]
\end{definition}

\section{Incentive Compatibility and Social Welfare Maximization} \label{sec:mechanisms}

Our primary goal is to design a mechanism that maximizes social welfare, defined as the sum of the advertisers' true values:
\[\totalWel(\allocs) \triangleq \sum_{\ad, \slot} \alloc[\ad, \slot]\cdot \wel[\ad, \slot].\]
Because payments are transfers from advertisers to the platform, they cancel when advertiser and platform utilities are summed.

However, a fundamental challenge is that the platform cannot observe the true value $\wel[\ad, \slot]$ derived from latent intents. It can only observe the proxy value $\welEs[\ad, \slot]$ derived from genres. This creates a misalignment between the \emph{proxy} objective optimized by the platform and the \emph{ground truth} value and utility experienced by advertisers (as shown in \Cref{fig:genre_model}).

In this section, we analyze how this misalignment affects incentive compatibility and social welfare. The proofs for all results in this section are deferred to \Cref{app:proofs}.

\subsection{Incentive Compatibility}

We first examine the incentives for advertisers. Since the platform designs the mechanism based on its own estimates, we define properties regarding the proxy setting versus the ground truth setting.

\begin{definition}[DSIC and IR in the proxy setting]
    We say a mechanism is dominant-strategy incentive-compatible (DSIC) in the proxy setting if truthful bidding maximizes proxy utility, i.e.,
    \[\utilEs(\valsEs[\ad], \bids[-\ad]) \geq \utilEs(\bids[\ad], \bids[-\ad]), \quad \forall \ad, \valsEs[\ad], \bids[\ad], \bids[-\ad]. \]
    We say a mechanism is individually rational (IR) in the proxy setting if truthful bidding yields non-negative utility, i.e.,
    \[\utilEs(\valsEs[\ad], \bids[-\ad]) \geq 0, \quad \forall \ad, \valsEs[\ad], \bids[-\ad]. \]
\end{definition}

However, advertisers may still deviate if their true utility $\util$ differs from $\utilEs$. Due to the platform's imperfect alignment with latent variables, we aim for \emph{approximate} guarantees in the ground truth setting relative to the upper bound on advertiser values $\valMax = \max_{\ad, \intent}{\val[\ad, \intent]}$. 

\begin{definition}[Approximate DSIC and IR in the ground truth setting]
    We say a mechanism is $\varepsilon$-DSIC in the ground-truth setting for $\varepsilon \geq 0$ if truthful bidding maximizes ground-truth utility up to an additive $\varepsilon\cdot \valMax$, i.e.,
    \[\util(\valsEs[\ad], \bids[-\ad]) + \varepsilon \cdot \valMax \geq \util(\bids[\ad], \bids[-\ad]), \quad \forall \ad, \valsEs[\ad], \bids[\ad], \bids[-\ad]. \]
    We say a mechanism is $\varepsilon$-IR in the ground-truth setting for $\varepsilon \geq 0$ if truthful bidding yields utility of at least $-\varepsilon\cdot \valMax$, i.e.,
    \[\util(\valsEs[\ad], \bids[-\ad]) \geq -\varepsilon\cdot \valMax, \quad \forall \ad, \valsEs[\ad], \bids[-\ad]. \]
\end{definition}

The following proposition establishes that if the approximation errors (from values on genres and coherence estimation) are bounded, a DSIC and IR mechanism in the proxy setting remains approximately DSIC and IR in the ground truth setting.

\begin{restatable}{proposition}{propDSIC}\label{prop:DSIC and IR transfer}
    If a mechanism is DSIC and IR in the proxy setting, it is $(2\varepsilon)$-DSIC and $\varepsilon$-IR in the ground truth setting, where $\varepsilon = \gapVal + \gapCoh$. 
\end{restatable}

\subsection{Social Welfare Maximization}

Because the ground-truth value $\wel[\ad, \slot]$ is unobservable to the platform, the platform can optimize only the \emph{proxy social welfare}:
\[\totalWelEs(\allocs) \triangleq \sum_{\ad, \slot} \alloc[\ad, \slot]\cdot \welEs[\ad, \slot]. \]
We now analyze the optimization of this proxy and derive a bound on the approximation error relative to the ground truth social welfare $\totalWel(\allocs)$.

As we have shown, for a DSIC and IR mechanism, deviating from truthful bidding can at most yield a utility gain of $2(\gapVal + \gapCoh)\cdot \valMax$ for each advertiser. 
Advertisers also do not know the ground-truth quantities $\val[\ad, \genre, \slot]$ and $\coh[\genre, \slot]$ at bidding time. We therefore evaluate welfare under truthful proxy bidding, the standard benchmark induced by the proxy DSIC mechanism.

\subsubsection{VCG Auction}

The Vickrey--Clarke--Groves (VCG) auction is known to satisfy DSIC and IR and to maximize social welfare under truthful bidding.
In our setting, after receiving all bids $\bids$ and computing the estimated coherence vector $\cohsEs[\slot]$ for each $\slot$, the platform computes an allocation $\allocsOPT$ that maximizes proxy social welfare $\totalWelEs(\allocs)$:
\begin{gather*}
    \allocsOPT \triangleq \argmax_{\allocs} \sum_{\ad, \slot} \alloc[\ad, \slot] \cdot \bids[\ad]^\top \cohsEs[\slot], \\
    \subjectTo \sum_{\slot} \alloc[\ad, \slot] \leq 1, \quad \forall \ad; \qquad
    \sum_{\ad} \alloc[\ad, \slot] \leq 1, \quad \forall \slot; \\
    \sum_{\ad, \slot} \alloc[\ad, \slot] = \numIns.
\end{gather*}

Each allocated advertiser pays the externality that its allocation imposes on the other advertisers:
\[\pay[\ad] = \max_{\allocs} \sum_{\ad' \neq \ad, \slot} \alloc[\ad', \slot] \cdot \bids[\ad']^\top \cohsEs[\slot] - \sum_{\ad' \neq \ad, \slot} \allocOPT[\ad', \slot] \cdot \bids[\ad']^\top \cohsEs[\slot].\]
In particular, when advertiser $\ad$ does not get allocated, $\pay[\ad] = 0$.

The main intuition behind the VCG auction is to align each advertiser's objective with the platform's objective---maximizing proxy social welfare---thereby guaranteeing DSIC and IR. To see this, compute each advertiser's proxy utility as
\begin{align*}
    \utilEs[\ad](\bids) &= \sum_{\slot} \allocOPT[\ad, \slot]\cdot \welEs[\ad, \slot] - \pay[\ad] \\
    &= \sum_{\slot} \allocOPT[\ad, \slot]\cdot \valsEs[\ad]^\top \cohsEs[\slot] + \sum_{\ad' \neq \ad, \slot} \allocOPT[\ad', \slot] \cdot \bids[\ad']^\top \cohsEs[\slot] \\
    &\enspace- \max_{\allocs} \sum_{\ad' \neq \ad, \slot} \alloc[\ad', \slot] \cdot \bids[\ad']^\top \cohsEs[\slot]. 
\end{align*}
The last term is independent of advertiser $\ad$'s bid; thus, its objective is to optimize
\[\sum_{\slot} \allocOPT[\ad, \slot]\cdot \valsEs[\ad]^\top \cohsEs[\slot] + \sum_{\ad' \neq \ad, \slot} \allocOPT[\ad', \slot] \cdot \bids[\ad']^\top \cohsEs[\slot].\]
Since $\allocsOPT$ optimizes $\sum_{\ad, \slot} \allocOPT[\ad, \slot]\cdot \bids[\ad]^\top \cohsEs[\slot]$, advertiser $\ad$'s optimal strategy is to bid $\bids[\ad] = \valsEs[\ad]$, and the corresponding utility is non-negative. 

\subsubsection{Social Welfare Gap}

Since the VCG auction that maximizes the proxy social welfare is DSIC and IR in the proxy setting, we can invoke \Cref{prop:DSIC and IR transfer} and derive that this mechanism is approximately DSIC and IR in the ground truth setting as well. 
However, even if we assume advertisers bid truthfully in this case, there remains a gap to the ground-truth social welfare $\totalWel(\allocs)$.
Below, we give a bound on the gap to show that the VCG mechanism behaves well.
\begin{restatable}{proposition}{propWelfare}\label{prop:welfare gap}
    If the platform uses the VCG auction to maximize proxy social welfare and advertisers bid truthfully, then the gap between the realized social welfare and the optimal ground-truth social welfare $\max_{\allocs} \totalWel(\allocs)$ is at most $2\numIns\cdot \varepsilon\cdot \valMax$, where $\varepsilon = \gapVal + \gapCoh$.
\end{restatable}

\subsection{Interpreting the Approximation Guarantees: The Role of Genres}

From \Cref{prop:DSIC and IR transfer,prop:welfare gap}, we establish that when the platform adopts the VCG mechanism with proxy values, the potential utility gain from strategic deviation is bounded by $2\varepsilon\cdot \valMax$ for each advertiser. Furthermore, assuming truthful bidding, the realized ground truth social welfare is suboptimal by at most $2\numIns\cdot \varepsilon\cdot \valMax$, with $\varepsilon = \gapVal + \gapCoh$.
Consequently, our framework provides conditional approximation guarantees that become informative when both $\gapVal$ and $\gapCoh$ are small.

While we discuss the estimation of coherence values $\coh[\genre, \slot]$ and the gap $\gapCoh$ later in \Cref{sec:coherence}, we focus here on drivers of a small valuation gap $\gapVal$. Recall that $\gapVal = \max_{\ad, \slot, \genre} \abs{\valEs[\ad,\genre] - \val[\ad,\genre,\slot]} / \valMax$. Intuitively, this gap is small if genres provide a good \emph{clustering} of intents, such that the distribution of user intents within a genre remains consistent across slot contexts.

To formalize this, let $\Dist[\genre, \slot]$ denote the conditional distribution of specific intents $\intent$ given a genre $\genre$ and slot $\slot$:
\[\prob[{\Dist[\genre, \slot]}]{\intent} \triangleq \frac{\prob{\intent \condition \slot}}{\prob{\genre \condition \slot}}, \quad \forall \intent \in \genre.\] 

We quantify the variation between these distributions across slots using total variation distance (TVD):
\[\TVD(\Dist, \Dist') \triangleq \frac{1}{2}\sum_{\intent\in\intentSpace}\abs{\prob[\Dist]{\intent}-\prob[\Dist']{\intent}}.\]
We now have the following result.
\begin{restatable}{proposition}{propTVD}\label{prop:small TVD leads to small gapVal}
    If $\TVD(\Dist[\genre, \slot], \Dist[\genre, \slot']) \leq \gapDis$ for every genre $\genre$ and pair of slots $\slot\neq \slot'$ satisfying $\prob{\genre\condition\slot}>0$ and $\prob{\genre\condition\slot'}>0$, then $\gapVal = \max_{\ad, \slot, \genre} \frac{\abs{\valEs[\ad,\genre] - \val[\ad,\genre,\slot]}}{\valMax} \leq \gapDis$.
\end{restatable}

\section{Coherence Measurement Implementation} \label{sec:coherence}

In this section, we describe implementations for estimating coherence $\coh[\genre, \slot] = \prob{\genre \condition \slot}$, the probability that a user's intent belongs to genre $\genre$ at slot $\slot$.

In a mature advertising ecosystem, this probability is typically estimated using extensive historical data (e.g., click-through rates or conversion logs). However, for a novel framework such as ad insertion in LLM responses, such production data may be unavailable, making it potentially intractable to directly measure this probability.

To bridge this gap, our framework introduces a two-step process: 
First, we compute a raw signal measuring the semantic alignment. We term this estimate the \emph{coherence signal}, denoted by $\scoEs[\genre, \slot] \in \reals$, which quantifies how well an ad of genre $\genre$ semantically aligns with the organic context at slot $\slot$. Second, we map this signal to a valid probability distribution. This process is often known as \emph{calibration} in machine learning \citep{guo2017calibration}. We formally define the calibration step as follows.

\begin{definition}[Calibration Function]
    We define the calibration function $\cali: \reals^{\numGen} \to \probSpace^{\numGen}$ as a mapping that converts the vector of coherence signals into a probability distribution over genres. For every slot $\slot$, the estimated coherence vector is given by:
    \[(\cohEs[\genre, \slot])_{\genre \in \genreSpace} = \cali((\scoEs[\genre, \slot])_{\genre \in \genreSpace}). \]
\end{definition}

In a production environment, $\cali$ is a learnable module (e.g., isotonic regression \citep{zadrozny2002transforming} or temperature-scaled softmax) that fits raw coherence signals to observed user interactions. We focus on deriving high-quality raw signals $\scoEs[\genre, \slot]$ and leave data-driven calibration to future deployment.

We now introduce two methods for computing the coherence signal, based on sentence embeddings and LLM-as-a-Judge, respectively.
In \Cref{sec:experiment-coherence}, we empirically compare these methods' alignment with human-annotated ratings.

\paragraph{Sentence embedding.} 
Sentence embedding models convert a sentence into a numerical vector that represents its semantic meaning. 
The cosine similarity between two embedding vectors can then be used to measure the semantic similarity between the sentences. 

Recall that the organic text is a list of sentences, denoted as $z = z_1z_2\dots z_M$. 
We define the embedding-based coherence signal $\scoEs[\genre, \slot]^{\operatorname{(embed)}}$ as the sum of the cosine similarities between the ad genre and its two adjacent sentences, minus the similarity between those adjacent sentences. The final term represents the cost of disrupting the original flow:
\begin{align*}
\scoEs[\genre, \slot]^{\operatorname{(embed)}} &= \operatorname{cos}(\operatorname{embed}(z_j), \operatorname{embed}(g)) \\
&\ + \operatorname{cos}(\operatorname{embed}(g), \operatorname{embed}(z_{j+1}))\\
&\ - \operatorname{cos}(\operatorname{embed}(z_j), \operatorname{embed}(z_{j+1})). 
\end{align*}

\paragraph{LLM-as-a-Judge.} 
Recent work has shown that LLMs can act as evaluators of natural-language quality, a paradigm known as ``LLM-as-a-Judge.'' Unlike embeddings, which capture local semantic overlap, this method uses an LLM's reasoning capabilities to evaluate rhetorical flow and contextual relevance.

For each candidate slot $j$, we present the user prompt, the organic response, the marked slot, and the list of ad genres to the LLM using the structured prompt in \Cref{app:prompts}.
For each genre $g$, the LLM outputs an estimated coherence signal $\scoEs[\genre, \slot]^{\operatorname{(judge)}} \in \{1,2,3,4,5\}$ and a brief rationale.

\section{Coherence Measurements' Alignment with Human Evaluation} \label{sec:experiment-coherence}

The two automated coherence measures assess how well ad genres integrate with surrounding organic content.
However, a critical question remains:
\begin{center}
    \textit{To what extent do these automated measurements align with human perceptions of ad--context coherence in LLM-generated content?}
\end{center}

To answer this question, we compare automated coherence estimates $\scoEs[\genre, \slot]$ with human ratings $\sco[\genre, \slot]$.
Our analysis spans a diverse set of organic prompts, ad genres, and insertion positions. 
To ensure both diversity and representativeness, we select prompts based on prior surveys of LLM user behavior, and choose ad genres grounded in real-world digital advertising expenditure statistics.

\paragraph{Experiment setup.} 
Specifically, we select seven diverse prompts that reflect typical user interactions with LLMs, informed by survey data from over 500 U.S. adults \citep{rainie2025close}: travel planning (reported by 34\% of participants), product suggestions (57\%), social-gathering planning (23\%), creative writing (36\%), term explanations (68\%), health information (39\%), and coding (23\%).
For each prompt, we generate an organic response using OpenAI's GPT-4o model.

We select 10 high-spend ad genres based on industry spending patterns in digital advertising~\citep{statista-digital-ad-verticals}. 
The selected genres represent sectors with the highest digital advertising expenditure.\footnote{To increase category granularity, we divide the retail sector into three sub-genres: packaged food, apparel, and electronics.} These genres are presented in alphabetical order in the experiment to mitigate ordering effects.

For each generated text, we identify potential ad insertion points primarily at paragraph boundaries, as these represent typical positions for display advertisements in digital content. 
For robustness testing, we also select several insertion points within paragraphs that may disrupt the flow of organic content, allowing us to evaluate the metrics across varying degrees of contextual disruption.

\paragraph{Participant recruitment.} 
We recruit 48 participants for our study in China, with 36 completing the entire experiment.\footnote{Other participants only complete the exploratory survey about online advertising in LLMs, as discussed in \Cref{sec:survey-results}.} 
According to our pilot study, each session lasts approximately 40 minutes, and participants receive 75 CNY (approximately US\$10.41) for their effort.

The demographic data of the participants show that all the participants have completed or are pursuing at least a college degree. 
For English proficiency, 77.1\% of participants reported an advanced level or higher, while the remainder reported an intermediate level. These characteristics suggest that participants could understand the English-language tasks, although they limit the generalizability of our results.
The detailed instructions for participants are presented in \Cref{app:experiment-instructions}. 

\paragraph{Coherence signal measurement setup.} 
For the sentence-embedding-based measurement, we use Qwen3 embedding models (0.6B, 4B, and 8B), deployed on an NVIDIA A100 80GB GPU.

For LLM-as-a-Judge, we use Qwen3 (8B and 32B), GPT-4o-mini, GPT-4o, GPT-5, DeepSeek-V3, and DeepSeek-R1 through hosted APIs. We include Qwen3-8B to compare embedding-based methods and LLM-as-a-Judge using models of the same size.

In \Cref{app:prompts}, we provide the prompt used for LLM-as-a-Judge coherence measurement.

\paragraph{Experiment procedure.} 
Across the 7 selected contexts, we have specified 31 possible ad insertion points in total. 
For each insertion point, participants rate the suitability of inserting each of the 10 ad genres on a scale from 1 to 5, following the instructions in \Cref{app:experiment-instructions}.
We define a single evaluation task as a unique (genre, position) pair. 
This results in 310 tasks in total.

\paragraph{Statistical metric.}
To represent overall human perception, we compute the average score from all participants for each task. 
We assess alignment by computing Spearman's rank correlation between the mean human score and each automated coherence measure over all 310 tasks.

\paragraph{Result 1: correlation.}

\begin{table*}[t]
\centering
\begingroup

\caption{Spearman correlations between automated coherence measures and mean human ratings across 310 genre--slot tasks (higher is better).}
\label{tab:metrics_summary}

\begin{minipage}[t]{\linewidth}
\centering
\par\textit{(a) Sentence-embedding-based methods.}
\vspace{0.3em}
\begin{tabular}{@{}lccc@{}}
\toprule
\textbf{Base Model} & \multicolumn{3}{c}{\textbf{Qwen3}} \\
\cmidrule(lr){2-4}
 & embedding-0.6B & embedding-4B & embedding-8B \\
\midrule
Spearman's $\rho$ & 0.2701 & 0.2838 & 0.3116 \\
\bottomrule
\end{tabular}
\end{minipage}

\begin{minipage}[t]{\linewidth}
\centering
\par\textit{(b) LLM-as-a-Judge methods.}
\vspace{0.3em}
\begin{tabular}{@{}lccccccc@{}}
\toprule
\textbf{Base Model} & \multicolumn{2}{c}{\textbf{Qwen3}} & \multicolumn{3}{c}{\textbf{GPT}} & \multicolumn{2}{c}{\textbf{DeepSeek}} \\
\cmidrule(lr){2-3}\cmidrule(lr){4-6}\cmidrule(lr){7-8}
 & 8B & 32B & 4o-mini & 4o & 5 & v3 & r1 \\
\midrule
Spearman's $\rho$ & 0.5532 & 0.5804 & 0.5856 & 0.6380 & 0.6637 & 0.6636 & 0.6650 \\
\bottomrule
\end{tabular}
\end{minipage}

\endgroup

\end{table*}

\Cref{tab:metrics_summary} presents the Spearman correlations between each automated measure and the human annotations. LLM-as-a-Judge correlates more strongly with human judgments than sentence embeddings, including when both use Qwen3-8B. Correlations generally increase from the smaller Qwen3 judges to the proprietary models; DeepSeek-R1 and GPT-5 both achieve approximately $0.66$. These results suggest that LLM-as-a-Judge may benefit from future improvements in foundation models.

\begin{figure}[ht]
    \centering
    \includegraphics[width=0.7\linewidth]{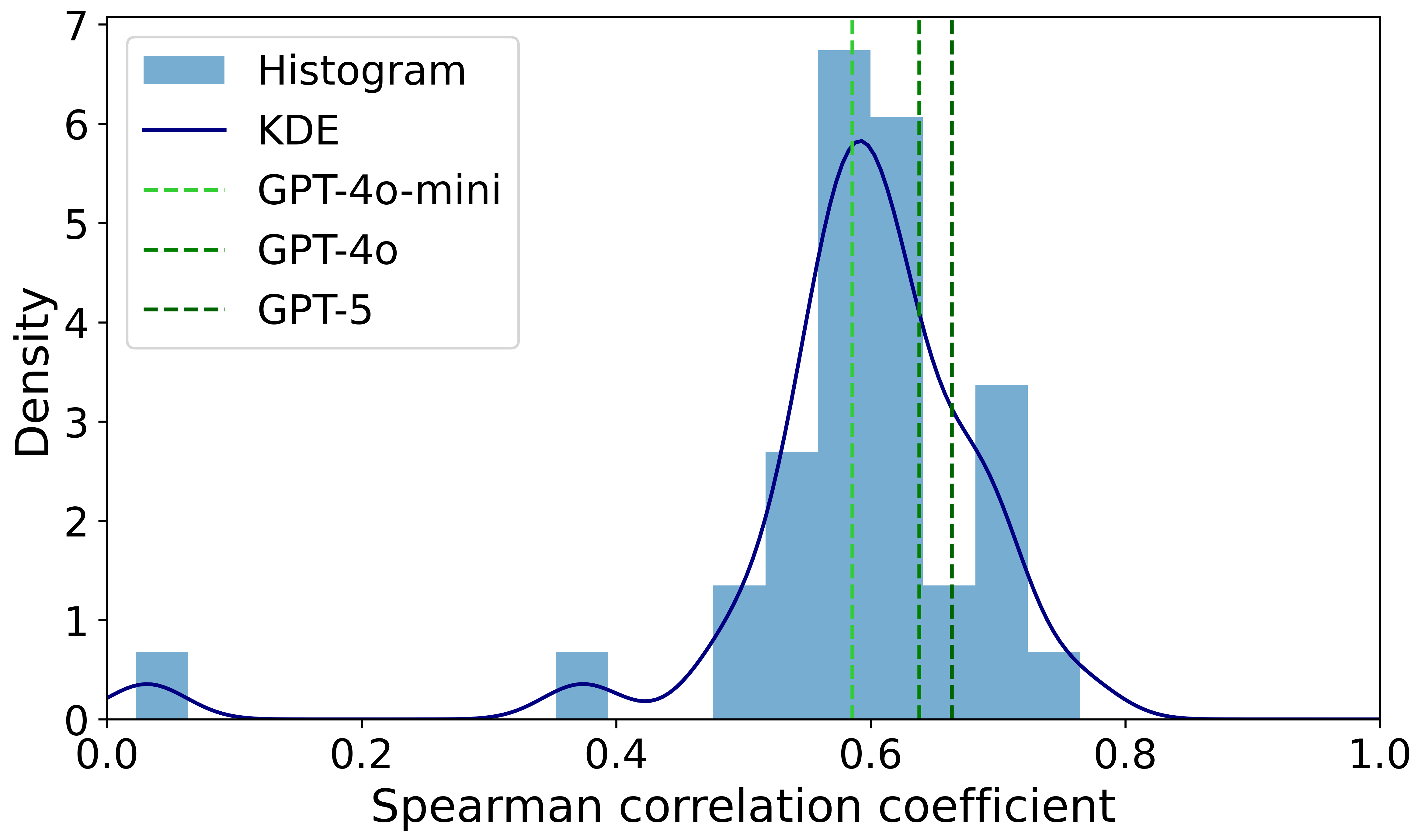}
    \caption{Distribution of leave-one-out individual-to-group Spearman correlations. Dashed lines show three GPT judges; GPT-5 exceeds 29 of 36 individual raters.}
    \label{fig:human-correlation-dist}
\end{figure}

\paragraph{Result 2: LLM-as-a-Judge aligns more closely with the group mean than most individual raters.}
To contextualize the correlations, we compare the LLM's performance with that of individual participants. For each participant, we compute the correlation between that participant's ratings and the mean ratings of all other participants. \Cref{fig:human-correlation-dist} compares the resulting distribution with the GPT-5 judge's correlation. The GPT-5 judge has a higher correlation with the leave-one-out group mean than 29 of 36 individual participants (80.6\%). One possible explanation is that modern LLMs are trained to reflect aggregate human preferences, whereas individual judgments contain idiosyncratic variation.

\paragraph{Result 3: The GPT-5 judge has a rater equivalence of approximately two.}
Following \citet{resnick2021survey}, we adopt the concept of \textit{rater equivalence}, which quantifies the minimum number of human raters needed to achieve the same expected correlation with the population mean as a given classifier. 
To estimate this, we compute the average correlation between the mean of $Q$ randomly sampled human participants (denoted as a committee) and the population mean (excluding the sampled committee to avoid bias). 
We compare these values with the correlation obtained by LLM-as-a-Judge.

\begin{figure}[ht]
    \centering
    \includegraphics[width=0.7\linewidth]{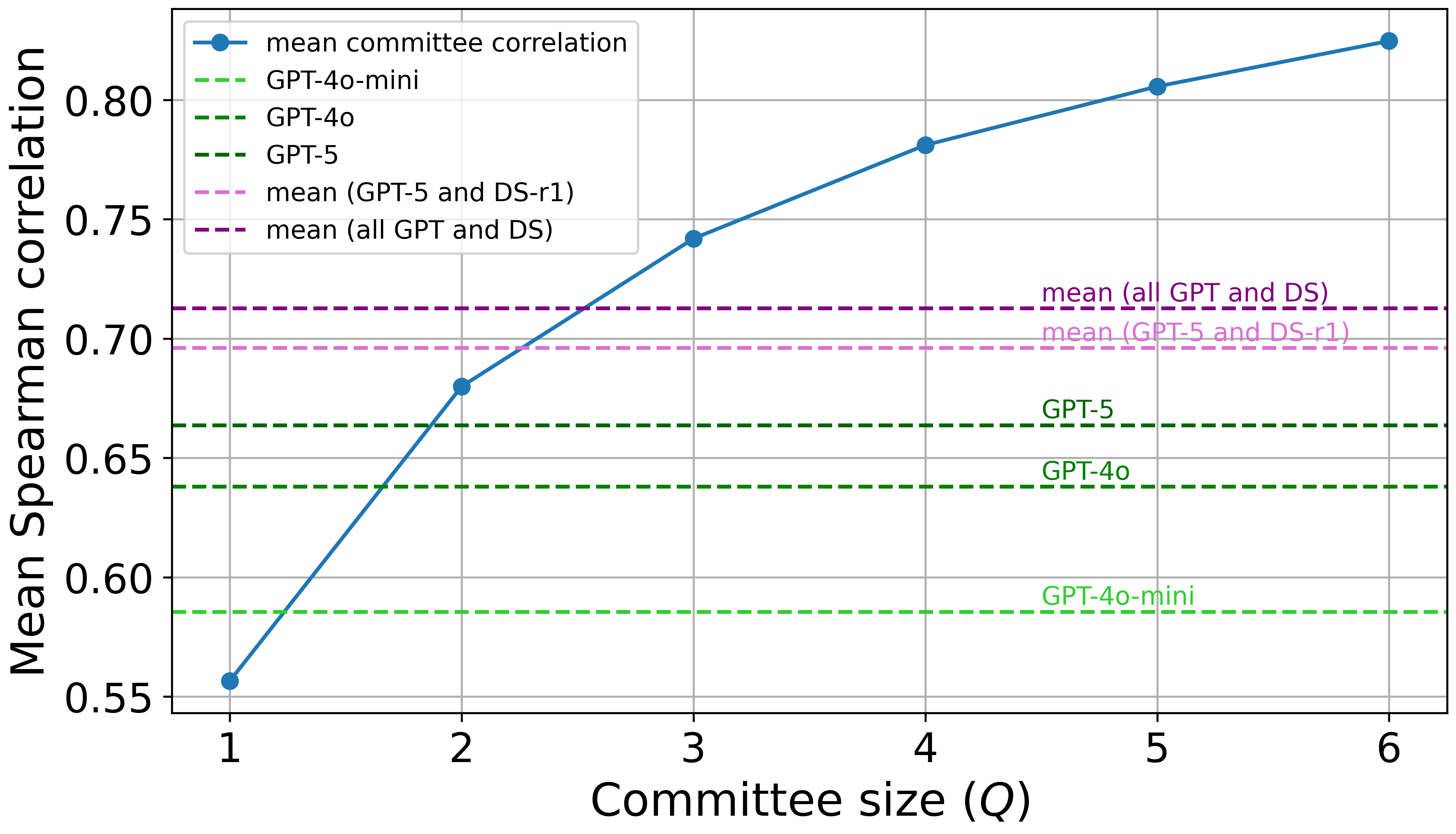}
    \caption{Average committee-to-group Spearman correlation by committee size. GPT-5 aligns with the group mean about as well as a two-person committee.}
    \label{fig:survey-eq}
\end{figure}

In \Cref{fig:survey-eq}, the GPT-5 judge has a rater equivalence of approximately two: its predictions align with the group mean about as well as the mean ratings of two randomly chosen participants.
Increasing the committee size substantially increases correlation with the population mean, whereas averaging multiple LLM-as-a-Judge instances yields only modest gains.
We hypothesize that this is because different LLMs might be less independent than human raters.

These experiments evaluate the ranking quality of the raw coherence signal, not its calibration as a probability distribution. As discussed in \Cref{sec:coherence}, calibrating the signal requires interaction data from a deployed system and remains future work.

\section{Prototypes} \label{sec:prototypes}

Putting all components together, we implement a prototype of the end-to-end workflow: it accepts genre-based bids and an organic response and produces an ad-augmented response with corresponding payments. Specifically, we use DeepSeek-R1 as the ``LLM-as-a-Judge'' coherence estimator and min--max normalize its scores to $[0,1]$.\footnote{This heuristic normalization is used only for the illustrative prototype. The approximation guarantees in \Cref{sec:mechanisms} require calibrated coherence probabilities, which we leave for future deployment.} We then apply the VCG mechanism to calculate allocations and payments.

\begin{figure}[ht]
    \centering
    \includegraphics[width=0.9\linewidth]{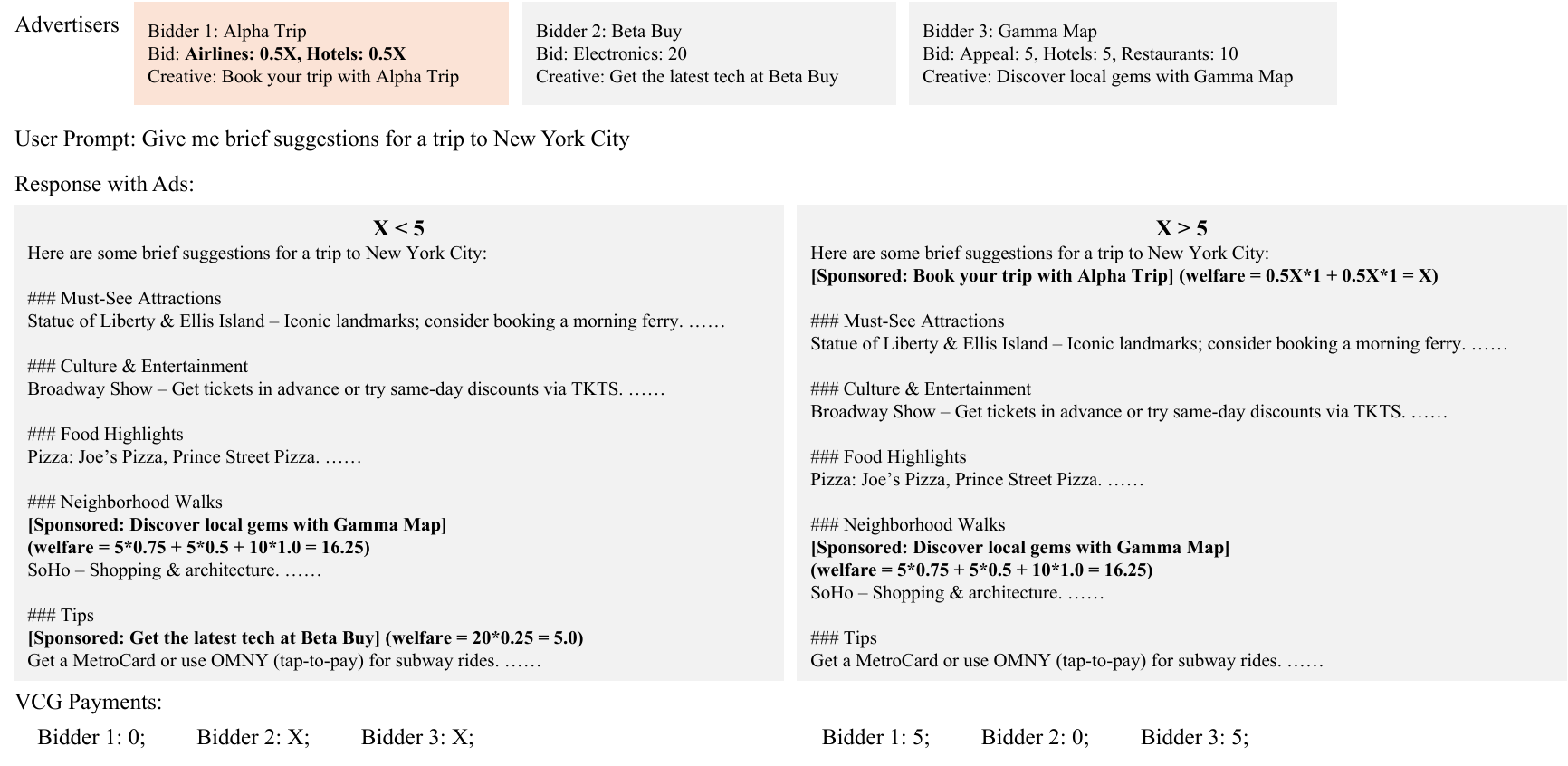}
    \caption{An ad-insertion example using the VCG mechanism ($\numIns = 2$) in a travel-advice scenario. Three advertisers participate: Alpha Trip (Airlines/Hotels), with a uniform bid of $0.5\bidV$ on both genres; Beta Buy, with a bid of $20$ on Electronics; and Gamma Map, with bids across Apparel, Hotels, and Restaurants. When $\bidV > 5$, Alpha Trip wins because of high coherence despite its lower raw bids. Payments reflect the welfare of the displaced bidder in each scenario.}
    \label{fig:demo}
\end{figure}

\Cref{fig:demo} illustrates a travel planning scenario in which we insert $\numIns = 2$ ads into an LLM-generated organic response. Notably, when Alpha Trip (advertiser 1) increases its bid $\bidV$ above 5, it is allocated an ad slot, despite its bid still being significantly less than Beta Buy's (advertiser 2) bid on Electronics (20). 
This occurs because Alpha Trip's targeted genres align better with the organic context: Hotels and Airlines both have the maximum coherence score of 1.0 at the beginning of the travel-advice response. Its resulting proxy welfare is $\bidV$, which exceeds Beta Buy's welfare of 5. Meanwhile, Gamma Map (advertiser 3) secures the ``Neighborhood Walks'' slot with the highest proxy welfare (16.25) by aggregating values across multiple high-coherence genres that align with the slot's context.

Finally, the calculated payments follow the VCG rule: winners pay the externality they impose on the system. For example, when Alpha Trip wins ($\bidV > 5$), it displaces Beta Buy (whose welfare is 5) and therefore pays 5.

\subsection{Computational Efficiency of VCG}

While VCG is often criticized for the computational cost of its maximum-weight matching at scale, the structure of LLM ad insertion mitigates this concern: the number of candidate slots $\numSlot$ is naturally bounded by the response length, and the number of inserted ads $\numIns$ is kept small to maintain user engagement.
We empirically evaluate this cost in synthetic experiments.
 
\paragraph{Setup.}
We use synthetic data to simulate a large-scale ad-insertion scenario. We fix the number of genres at $\numGen = 100$ and the number of inserted ads at $\numIns = 5$. We vary the number of candidate slots $\numSlot \in \{20, 50, 100\}$ and advertisers $\numAd \in \{10^3, 10^4, 10^5\}$. For each pair $(\numAd, \numSlot)$, we run 100 independent simulation trials.

We solve the matching problem using SciPy's implementation of the Jonker--Volgenant (JV) algorithm \citep{scipy-linear-sum-assignment}. All experiments run on a consumer-grade laptop (2020 Apple MacBook Air with an M1 chip).

\paragraph{Synthetic data generation.}
For advertiser bids, we assume a sparse interest model where each advertiser bids on a limited subset of genres. The size of this subset is determined by a Poisson distribution ($\lambda=2$), reflecting that most advertisers target only a few specific verticals. For these selected genres, the bid values are drawn uniformly from the interval $[0,1]$, while valuations for non-selected genres are set to zero.

We also generate a coherence matrix representing the alignment between every candidate slot and ad genre. We draw coherence scores uniformly from $[0,1]$ and randomly set half of the entries to zero to simulate the common case in which many genres are irrelevant to a slot. We ensure that every slot has non-zero coherence.

\paragraph{Results.}
\Cref{tab:prototype} presents the mean end-to-end execution time for each configuration, measured from processing the bids and coherence matrix to producing allocations and payments. Runtime scales approximately linearly with the numbers of advertisers $\numAd$ and slots $\numSlot$. In the largest scenario ($\numAd = 10^5$, $\numSlot = 100$), VCG clears in approximately $1.25$ seconds on the test laptop.

\begin{table}[ht]
  \centering
  \caption{Mean VCG execution time in seconds.}
  \label{tab:prototype}
  \begin{tabular}{cccc}
      \toprule
       & $\numSlot = 20$ & $\numSlot = 50$ & $\numSlot = 100$ \\
      \midrule
    $\numAd = 10^3$  & 0.0030 & 0.0064 & 0.0121 \\
    $\numAd = 10^4$  & 0.0293 & 0.0659 & 0.1256 \\
    $\numAd = 10^5$  & 0.3110 & 0.6946 & 1.2475 \\
    \bottomrule
  \end{tabular}
\end{table}


Even in the largest scenario tested ($\numAd = 10^5$ advertisers, $\numSlot = 100$ slots, $\numIns = 5$, and $\numGen = 100$ genres), VCG clears in approximately $1.25$ seconds on a consumer-grade laptop, with execution time scaling approximately linearly in both $\numAd$ and $\numSlot$.
Because LLM inference often takes seconds, these results suggest that the auction itself can be computationally feasible for real-time ad insertion.

\section{Related Work}

\paragraph{Ad insertion in LLMs.}
Recent scholarship has focused on integrating ads into LLM responses. A position paper by \citet{feizi2023online} outlines critical components in this domain: the modification module (how to display ads), the bidding module (what to bid for), the prediction module (estimating CTR), and the auction module (mechanism design).

On the question of ``what to bid for,'' existing work falls into two broad categories: bidding at the \emph{token or segment level} \citep{duetting2024mechanism,hajiaghayi2024ad,zhao2025llm,han2026mechanism} and bidding at the \emph{query or response level} \citep{soumalias2024truthful,dubey2024auctions,mordo2024sponsored,balseiro2025position}.
As discussed in \Cref{sec:intro}, both approaches raise practical concerns, including computational cost, hallucination risk, and privacy leakage. Our framework instead introduces bidding on \emph{genres} to mitigate these tensions.

Other recent mechanism-design papers, including \citet{banchio2025ads} and \citet{bergemann2025data}, propose mechanisms that may apply to ad auctions in LLM responses but focus primarily on the abstract mechanism-design problem.
In contrast, we provide an end-to-end framework that connects bidding, contextual-coherence estimation, allocation, and payment calculation.

\paragraph{Evaluation metrics for natural language generation (NLG).}
Natural language generation tasks such as machine translation and summarization have traditionally relied on automatic metrics that compare a model’s output to human-written reference texts. Classic metrics like BLEU~\citep{papineni2002bleu} and ROUGE~\citep{lin2004rouge} measure n-gram overlap with reference translations or summaries. In recent years, LLM-based metrics have been proposed to capture deeper semantic quality. For example, BERTScore~\citep{zhang2019bertscore} uses contextual embedding similarity. BARTScore~\citep{yuan2021bartscore}, GPTScore~\citep{fu2024gptscore}, and GEM~\citep{xu2024benchmarking,lu2024eliciting} use pre-trained generative models to score how well a candidate can reproduce the reference or vice versa. 

However, inserting advertisements into an LLM-generated response is a new task. We still care about the overall language quality of the augmented response, but the primary evaluation focus is on contextual coherence. Recent work on LLM-as-a-judge evaluation~\citep{bai2024benchmarking,ke2023critiquellm,zheng2023judging} supports using LLMs' reasoning ability to rate outputs according to the given criteria, which is promising for our needs. In addition, we propose a sentence-embedding-based method that uses cosine similarity between ads and their surrounding content as an additional approach for evaluating contextual coherence.

\section{Conclusion and Discussion}\label{sec:conclusion}

We present a framework for inserting ads into LLM responses while addressing advertiser welfare, contextual coherence, and computational efficiency. The framework uses two layers of decoupling. First, it separates organic response generation from ad insertion, facilitating ad pre-screening and explicit disclosure. Second, it introduces ``genres'' that separate bidding from real-time response generation, reducing the privacy and latency costs of sharing full user prompts or responses. Genres thus provide a stable, biddable interface that functions as a counterpart to keywords in search advertising.


Our study has certain limitations that suggest directions for future work. 

First, our framework relies on language models to provide accurate coherence measurements. While our LLM-as-a-Judge method aligns with human annotations, further refinements could improve accuracy and efficiency. Future work may explore task-aware embeddings \citep{liu2018task} tailored specifically to ad--context fit or fine-tune the evaluator via RLHF \citep{christiano2017deep} to better capture user intent for ad insertion.
Additionally, mapping the raw coherence signals to calibrated probabilities using production data remains a crucial step for real-world deployment.
Advances in embedding learning, AI alignment, and generative recommendation systems \citep{he2025plum} will likely offer useful tools in the future.

Second, using genres mediates the trade-off between efficiency and contextual coherence. Our current model assumes a fixed partition of the user-intent space. Future work could explore adaptive partitioning or hierarchical taxonomies (e.g., \emph{Automotive} $\rightarrow$ \emph{Cars} $\rightarrow$ \emph{Sports cars}) under explicit runtime constraints to determine the genre granularity dynamically, minimizing the valuation error $\gapVal$ while maintaining scalability.

Third, our human evaluation uses 36 participants from a relatively homogeneous, highly educated population in China and only seven prompt contexts. Broader and more diverse participant populations, prompts, and commercial settings are needed before generalizing the observed correlations.



Finally, ad-supported LLMs may raise a platform-incentive concern: when revenue is tied to impressions or clicks, providers may optimize for engagement (e.g., longer interactions that create more opportunities to show ads) \citep{anwar2025recommendation}, potentially distorting the ``organic'' response. Future work should examine mechanisms that better align platform objectives with user trust and preferences \citep{kleinberg2024challenge}.

\bibliographystyle{plainnat}
\bibliography{reference}

@inproceedings{anwar2025recommendation,
  title={Recommendation and temptation},
  author={Anwar, Md Sanzeed and Dhillon, Paramveer S and Schoenebeck, Grant},
  booktitle={Proceedings of the Nineteenth ACM Conference on Recommender Systems},
  pages={422--431},
  year={2025}
}

@article{kleinberg2024challenge,
  title={The challenge of understanding what users want: Inconsistent preferences and engagement optimization},
  author={Kleinberg, Jon and Mullainathan, Sendhil and Raghavan, Manish},
  journal={Management science},
  volume={70},
  number={9},
  pages={6336--6355},
  year={2024},
  publisher={INFORMS}
}

@misc{iab_taxonomy,
  author       = {{IAB Tech Lab}},
  title        = {Content Taxonomy},
  howpublished = {\url{https://iabtechlab.com/standards/content-taxonomy/}},
  year         = {2024},
  note         = {Accessed: 2025-12-02}
}

@article{he2025plum,
  title={{PLUM}: Adapting Pre-trained Language Models for Industrial-scale Generative Recommendations},
  author={He, Ruining and Heldt, Lukasz and Hong, Lichan and Keshavan, Raghunandan and Mao, Shifan and Mehta, Nikhil and Su, Zhengyang and Tsai, Alicia and Wang, Yueqi},
  journal={arXiv preprint arXiv:2510.07784},
  year={2025}
}

@inproceedings{guo2017calibration,
  title={On Calibration of Modern Neural Networks},
  author={Guo, Chuan and Pleiss, Geoff and Sun, Yu and Weinberger, Kilian Q},
  booktitle={Proceedings of the 34th International Conference on Machine Learning},
  pages={1321--1330},
  year={2017},
  organization={PMLR}
}

@inproceedings{zadrozny2002transforming,
  title={Transforming classifier scores into accurate multiclass probability estimates},
  author={Zadrozny, Bianca and Elkan, Charles},
  booktitle={Proceedings of the Eighth ACM SIGKDD International Conference on Knowledge Discovery and Data Mining},
  pages={694--699},
  year={2002}
}

@article{broder2002taxonomy,
  title={A taxonomy of web search},
  author={Broder, Andrei},
  journal={ACM SIGIR Forum},
  volume={36},
  number={2},
  pages={3--10},
  year={2002},
  organization={ACM New York, NY, USA}
}

@inproceedings{ribeiro2005impedance,
  title={Impedance coupling in content-targeted advertising},
  author={Ribeiro-Neto, Berthier and Cristo, Marco and Golgher, Paulo B and Silva de Moura, Edleno},
  booktitle={Proceedings of the 28th annual international ACM SIGIR conference on Research and development in information retrieval},
  pages={496--503},
  year={2005}
}

@inproceedings{grbovic2016scalable,
  title={Scalable semantic matching of queries to ads in sponsored search advertising},
  author={Grbovic, Mihajlo and Djuric, Nemanja and Radosavljevic, Vladan and Silvestri, Fabrizio and Bhamidipati, Narayan},
  booktitle={Proceedings of the 39th International ACM SIGIR conference on Research and Development in Information Retrieval},
  pages={375--384},
  year={2016}
}

@inproceedings{liu2018task,
  title={Task-oriented word embedding for text classification},
  author={Liu, Qian and Huang, He-Yan and Gao, Yang and Wei, Xiaochi and Tian, Yuxin and Liu, Luyang},
  booktitle={Proceedings of the 27th International Conference on Computational Linguistics},
  pages={2023--2032},
  year={2018}
}

@article{christiano2017deep,
  title={Deep reinforcement learning from human preferences},
  author={Christiano, Paul F and Leike, Jan and Brown, Tom and Martic, Miljan and Legg, Shane and Amodei, Dario},
  journal={Advances in Neural Information Processing Systems},
  volume={30},
  pages={4302--4310},
  year={2017}
}

@misc{openai2026scaling,
  author       = {{OpenAI}},
  title        = {Scaling {AI} for Everyone},
  year         = {2026},
  howpublished = {\url{https://openai.com/index/scaling-ai-for-everyone/}}
}

@techreport{alphabet2024annualreport,
  author      = {{Alphabet Inc.}},
  title       = {2024 Annual Report},
  institution = {Alphabet Inc.},
  year        = {2024},
  url         = {https://abc.xyz/assets/99/21/46cafdba41089a12a2d86ea47d44/goog026-annualreport2024-web.pdf}
}

@misc{idc2024worldwide,
  author = {IDC},
  title = {Worldwide Spending on Artificial Intelligence Forecast to Reach \$632 Billion in 2028, According to a New IDC Spending Guide},
  year = {2024},
  howpublished = {\url{https://my.idc.com/getdoc.jsp?containerId=prUS52530724}}
}

@misc{ftc2015enforcement,
  author = {FTC},
  title = {Enforcement Policy Statement on Deceptively Formatted Advertisements},
  year = {2015},
  howpublished = {\url{https://www.ftc.gov/system/files/documents/public_statements/896923/151222deceptiveenforcement.pdf}}
}

@misc{statista-digital-ad-verticals,
  author       = {Statista},
  title        = {The Top Ad Spending Verticals in the {U.S.}},
  year         = {2025},
  howpublished = {\url{https://www.statista.com/chart/19241/top-10-digital-ad-spending-verticals/}}
}

@misc{scipy-linear-sum-assignment,
  author       = {SciPy},
  title        = {linear\_sum\_assignment — SciPy v1.16.2 Manual},
  year         = {2025},
  howpublished = {\url{https://docs.scipy.org/doc/scipy/reference/generated/scipy.optimize.linear_sum_assignment.html}}
}

@article{rainie2025close,
  title={Close encounters of the AI kind: The increasingly human-like way people are engaging with language models},
  author={Rainie, Lee},
  journal={https://imaginingthedigitalfuture.org/wp-content/uploads/2025/03/ITDF-LLM-User-Report-3-12-25.pdf},
  year={2025}
}

@article{resnick2021survey,
  title={Survey equivalence: A procedure for measuring classifier accuracy against human labels},
  author={Resnick, Paul and Kong, Yuqing and Schoenebeck, Grant and Weninger, Tim},
  journal={arXiv preprint arXiv:2106.01254},
  year={2021}
}

@inproceedings{lu2024eliciting,
  title={Eliciting informative text evaluations with large language models},
  author={Lu, Yuxuan and Xu, Shengwei and Zhang, Yichi and Kong, Yuqing and Schoenebeck, Grant},
  booktitle={Proceedings of the 25th ACM Conference on Economics and Computation},
  pages={582--612},
  year={2024}
}

@inproceedings{xu2024benchmarking,
  title={Benchmarking {LLMs}' Judgments with No Gold Standard},
  author={Xu, Shengwei and Lu, Yuxuan and Schoenebeck, Grant and Kong, Yuqing},
  booktitle={The Thirteenth International Conference on Learning Representations},
  year={2025}
}

@inproceedings{duetting2024mechanism,
  title={Mechanism design for large language models},
  author={Duetting, Paul and Mirrokni, Vahab and Paes Leme, Renato and Xu, Haifeng and Zuo, Song},
  booktitle={Proceedings of the ACM Web Conference 2024},
  pages={144--155},
  year={2024}
}

@article{soumalias2024truthful,
  title={Truthful aggregation of {LLMs} with an application to online advertising},
  author={Soumalias, Ermis and Curry, Michael J and Seuken, Sven},
  journal={arXiv preprint arXiv:2405.05905},
  year={2024}
}

@article{hajiaghayi2024ad,
  title={Ad auctions for {LLMs} via retrieval augmented generation},
  author={Hajiaghayi, MohammadTaghi and Lahaie, S{\'e}bastien and Rezaei, Keivan and Shin, Suho},
  journal={Advances in Neural Information Processing Systems},
  volume={37},
  pages={18445--18480},
  year={2024}
}

@inproceedings{dubey2024auctions,
  title={Auctions with {LLM} summaries},
  author={Dubey, Avinava and Feng, Zhe and Kidambi, Rahul and Mehta, Aranyak and Wang, Di},
  booktitle={Proceedings of the 30th ACM SIGKDD Conference on Knowledge Discovery and Data Mining},
  pages={713--722},
  year={2024}
}

@inproceedings{mordo2024sponsored,
  title={Sponsored question answering},
  author={Mordo, Tommy and Tennenholtz, Moshe and Kurland, Oren},
  booktitle={Proceedings of the 2024 ACM SIGIR International Conference on Theory of Information Retrieval},
  pages={167--173},
  year={2024}
}

@article{balseiro2025position,
  title={Position Auctions in AI-Generated Content},
  author={Balseiro, Santiago and Bhawalkar, Kshipra and Deng, Yuan and Feng, Zhe and Mao, Jieming and Mehta, Aranyak and Mirrokni, Vahab and Leme, Renato Paes and Wang, Di and Zuo, Song},
  journal={arXiv preprint arXiv:2506.03309},
  year={2025}
}

@article{feizi2023online,
  title={Online advertisements with {LLMs}: Opportunities and challenges},
  author={Feizi, Soheil and Hajiaghayi, MohammadTaghi and Rezaei, Keivan and Shin, Suho},
  journal={arXiv preprint arXiv:2311.07601},
  year={2023}
}

@inproceedings{devlin2019bert,
  title={Bert: Pre-training of deep bidirectional transformers for language understanding},
  author={Devlin, Jacob and Chang, Ming-Wei and Lee, Kenton and Toutanova, Kristina},
  booktitle={Proceedings of the 2019 Conference of the North American Chapter of the Association for Computational Linguistics: Human Language Technologies, Volume 1 (Long and Short Papers)},
  pages={4171--4186},
  year={2019}
}

@article{yuan2021bartscore,
  title={Bartscore: Evaluating generated text as text generation},
  author={Yuan, Weizhe and Neubig, Graham and Liu, Pengfei},
  journal={Advances in Neural Information Processing Systems},
  volume={34},
  pages={27263--27277},
  year={2021}
}

@inproceedings{fu2024gptscore,
  title={Gptscore: Evaluate as you desire},
  author={Fu, Jinlan and Ng, See Kiong and Jiang, Zhengbao and Liu, Pengfei},
  booktitle={Proceedings of the 2024 Conference of the North American Chapter of the Association for Computational Linguistics: Human Language Technologies (Volume 1: Long Papers)},
  pages={6556--6576},
  year={2024}
}

@article{bai2024benchmarking,
  title={Benchmarking foundation models with language-model-as-an-examiner},
  author={Bai, Yushi and Ying, Jiahao and Cao, Yixin and Lv, Xin and He, Yuze and Wang, Xiaozhi and Yu, Jifan and Zeng, Kaisheng and Xiao, Yijia and Lyu, Haozhe and Zhang, Jiayin and Li, Juanzi and Hou, Lei},
  journal={Advances in Neural Information Processing Systems},
  volume={36},
  pages={78142--78167},
  year={2024}
}

@inproceedings{zhang2019bertscore,
  title={{BERTScore}: Evaluating Text Generation with BERT},
  author={Zhang, Tianyi and Kishore, Varsha and Wu, Felix and Weinberger, Kilian Q and Artzi, Yoav},
  booktitle={International Conference on Learning Representations},
  year={2020}
}

@inproceedings{ke2023critiquellm,
  title={{CritiqueLLM}: Towards an Informative Critique Generation Model for Evaluation of Large Language Model Generation},
  author={Ke, Pei and Wen, Bosi and Feng, Zhuoer and Liu, Xiao and Lei, Xuanyu and Cheng, Jiale and Wang, Shengyuan and Zeng, Aohan and Dong, Yuxiao and Wang, Hongning and Tang, Jie and Huang, Minlie},
  booktitle={Proceedings of the 2023 Conference on Empirical Methods in Natural Language Processing},
  pages={9004--9017},
  year={2023}
}

@article{zheng2023judging,
  title = {Judging {LLM-as-a-Judge} with MT-Bench and Chatbot Arena},
  author = {Zheng, Lianmin and Chiang, Wei-Lin and Sheng, Ying and Zhuang, Siyuan and Wu, Zhanghao and Zhuang, Yonghao and Lin, Zi and Li, Zhuohan and Li, Dacheng and Xing, Eric and Zhang, Hao and Gonzalez, Joseph E and Stoica, Ion},
  journal = {Advances in Neural Information Processing Systems},
  volume = {36},
  pages = {46595--46623},
  year = {2023}
}

@inproceedings{lin2004rouge,
  title={Rouge: A package for automatic evaluation of summaries},
  author={Lin, Chin-Yew},
  booktitle={Text summarization branches out},
  pages={74--81},
  year={2004}
}

@inproceedings{papineni2002bleu,
  title={Bleu: a method for automatic evaluation of machine translation},
  author={Papineni, Kishore and Roukos, Salim and Ward, Todd and Zhu, Wei-Jing},
  booktitle={Proceedings of the 40th annual meeting of the Association for Computational Linguistics},
  pages={311--318},
  year={2002}
}

@inproceedings{zelch2024user,
  title={A User Study on the Acceptance of Native Advertising in Generative {IR}},
  author={Zelch, Ines and Hagen, Matthias and Potthast, Martin},
  booktitle={Proceedings of the 2024 Conference on Human Information Interaction and Retrieval},
  pages={142--152},
  year={2024}
}

@inproceedings{banchio2025ads,
  title={Ads in Conversations},
  author={Banchio, Martino and Mehta, Aranyak and Perlroth, Andres},
  booktitle={Proceedings of the 26th ACM Conference on Economics and Computation},
  pages={350--350},
  year={2025}
}

@inproceedings{bergemann2025data,
  title={Data-Driven Mechanism Design: Jointly Eliciting Preferences and Information},
  author={Bergemann, Dirk and Bojko, Marek and Duetting, Paul and Paes Leme, Renato and Xu, Haifeng and Zuo, Song},
  booktitle={Proceedings of the 26th ACM Conference on Economics and Computation},
  pages={507--507},
  year={2025}
}

@article{zhao2025llm,
  title={LLM-Auction: Generative Auction towards LLM-Native Advertising},
  author={Zhao, Chujie and Hu, Qun and Song, Shiping and Chen, Dagui and Zhu, Han and Xu, Jian and Zheng, Bo},
  journal={arXiv preprint arXiv:2512.10551},
  year={2025}
}

@article{han2026mechanism,
  title={Mechanism Design for Quality-Preserving LLM Advertising},
  author={Han, Jiale and Dai, Xiaowu},
  journal={arXiv preprint arXiv:2605.10964},
  year={2026}
}

\appendix
\section{Connection to Low-Rank Approximation of the Welfare Matrix}
\label{app:low-rank}

Combining the genre-based valuation proxy from \Cref{subsec:valuation} and the contextual coherence, we can interpret our framework as a low-rank approximation of the welfare matrix, as illustrated in \Cref{fig:welfare_matrix_fac}.

To formalize this, let $\cohsEs[\slot] \triangleq (\cohEs[\genre, \slot])_{\genre \in \{1, \dots, \numGen\}}$ denote the estimated coherence vector for slot $\slot$, and let $\valsEs[\ad] \triangleq (\valEs[\ad, \genre])_{\genre \in \{1, \dots, \numGen\}}$ denote the genre-based valuation vector for advertiser $\ad$. The estimated value for advertiser $\ad$ at slot $\slot$ is the inner product $\welEs[\ad, \slot] = \valsEs[\ad]^\top \cohsEs[\slot]$. Let $\welsMat_{\numAd \times \numSlot}\triangleq (\wel[\ad, \slot])$ denote the underlying value matrix, whose rows represent advertisers and whose columns represent response contexts (slots). Our framework approximates $\welsMat$ with the product of two lower-dimensional matrices:
\begin{enumerate}
    \item A static valuation matrix $\valsEsMat_{\numGen \times \numAd} \triangleq (\valsEs[\ad])_{\ad \in \{1, \dots, \numAd\}}$ (genres $\times$ advertisers), which can be elicited by offline bidding and held fixed during LLM inference.
    \item A dynamic coherence matrix $\cohsEsMat_{\numGen \times \numSlot} \triangleq (\cohsEs[\slot])_{\slot \in \{1, \dots, \numSlot\}}$ (genres $\times$ current slots), which the platform updates during real-time inference.
\end{enumerate}
The resulting approximation is $\welsMat \approx \valsEsMat^\top\cohsEsMat$, whose rank is at most the number of genres $\numGen$. By limiting $\numGen$ to a finite number (e.g., the size of the IAB taxonomy), we reduce the bidding dimensionality from the intractable context space to the tractable genre space.

\section{Omitted Proofs}

\label{app:proofs}

\propDSIC*

\begin{proof}[Proof of \Cref{prop:DSIC and IR transfer}]
    For any bids $\bids$, we bound the utility gap between two settings as follows:
    \begin{align*}
        &\mathrel{\phantom{=}} \abs{\utilEs(\bids) - \util(\bids)} \\
        &= \abs{\sum_{\slot} \alloc[\ad, \slot]\cdot \inParentheses{\valsEs[\ad]^\top \cohsEs[\slot] - \vals[\ad, \slot]^\top \cohs[\slot]}} \tag{where $\vals[\ad, \slot] \triangleq (\val[\ad, \genre, \slot])^\top_{\genre \in \{1, 2, \cdots, \numGen\}}$ } \\
        &= \abs{\sum_{\slot} \alloc[\ad, \slot]\cdot \inBrackets{(\valsEs[\ad] - \vals[\ad, \slot])^\top \cohsEs[\slot] +  \vals[\ad, \slot]^\top (\cohsEs[\slot] - \cohs[\slot])}} \\
        &\leq \sum_{\slot} \alloc[\ad, \slot] \cdot \inBrackets{ \norm{\infty}{\valsEs[\ad] - \vals[\ad, \slot]} \cdot \norm{1}{\cohsEs[\slot]} + \norm{\infty}{\vals[\ad, \slot]} \norm{1}{\cohsEs[\slot] - \cohs[\slot]}} \tag{H\"older's inequality} \\
        &\leq \sum_{\slot} \alloc[\ad, \slot] \cdot \inBrackets{ (\gapVal \cdot \valMax) \cdot 1 + \valMax \cdot \gapCoh } \\
        &\leq (\gapVal + \gapCoh)\cdot \valMax.
    \end{align*}

    Therefore, since the mechanism is DSIC in the proxy setting, we have for any $\ad, \valsEs[\ad], \bids[\ad], \bids[-\ad]$, 
    \begin{align*}
        \util(\bids[\ad], \bids[-\ad]) &\leq \utilEs(\bids[\ad], \bids[-\ad]) + (\gapVal + \gapCoh)\cdot \valMax \\
        &\leq \utilEs(\valsEs[\ad], \bids[-\ad]) + (\gapVal + \gapCoh)\cdot \valMax \\
        &\leq \util(\valsEs[\ad], \bids[-\ad]) + 2(\gapVal + \gapCoh)\cdot \valMax.
    \end{align*}
    This implies that the mechanism is $(2\varepsilon)$-DSIC in the ground truth setting for $\varepsilon = \gapVal + \gapCoh$. Similarly, 
    \begin{align*}
        \util(\valsEs[\ad], \bids[-\ad]) \geq \utilEs(\valsEs[\ad], \bids[-\ad]) - (\gapVal + \gapCoh)\cdot \valMax \geq - (\gapVal + \gapCoh)\cdot \valMax,
    \end{align*}
    and the mechanism is $\varepsilon$-IR in the ground truth setting. 
\end{proof}

\propWelfare*

\begin{proof}[Proof of \Cref{prop:welfare gap}]
    By the same norm bound used in the proof of \Cref{prop:DSIC and IR transfer}, $\abs{\welEs[\ad, \slot] - \wel[\ad, \slot]} \leq \varepsilon\cdot \valMax$ for every advertiser--slot pair. Thus, we have
    \begin{align*}
        \sum_{\ad, \slot} \allocOPT[\ad, \slot]\cdot \welEs[\ad, \slot] - \sum_{\ad, \slot} \allocOPT[\ad, \slot]\cdot \wel[\ad, \slot] \leq \sum_{\ad, \slot} \allocOPT[\ad, \slot]\cdot \varepsilon\cdot \valMax
        = \numIns\cdot \varepsilon\cdot \valMax. 
    \end{align*}
    This also implies that $\max_{\allocs} \totalWel(\allocs) \leq \max_{\allocs} \totalWelEs(\allocs) + \numIns\cdot \varepsilon\cdot \valMax$. Putting them together induces the result. 
\end{proof}

\propTVD*

\begin{proof}[Proof of \Cref{prop:small TVD leads to small gapVal}]
    Recall that 
    \begin{align*}
        \val[\ad,\genre,\slot] = \sum_{\intent \in \genre} \val[\ad,\intent] \cdot \prob[{\Dist[\genre, \slot]}]{\intent}.
    \end{align*}
    Using the property that the difference in expectations is bounded by the TVD scaled by the maximum value, and given $\TVD(\Dist[\genre, \slot], \Dist[\genre, \slot']) \leq \gapDis$, we have for any $\slot \neq \slot'$:
    \[ 
    \abs{\val[\ad,\genre,\slot] - \val[\ad,\genre,\slot']} \leq \valMax \cdot \TVD(\Dist[\genre, \slot], \Dist[\genre, \slot']) \leq \gapDis \cdot \valMax.
    \]
    
    Since $\valEs[\ad,\genre]$ is the average value over the reference distribution of contexts in which genre $\genre$ has positive probability, the distance between any single $\val[\ad, \genre, \slot]$ and the mean $\valEs[\ad, \genre]$ cannot exceed the maximum pairwise distance between $\val[\ad, \genre, \slot]$ and $\val[\ad, \genre, \slot']$:
    \[ \abs{\valEs[\ad, \genre] - \val[\ad, \genre, \slot]} \leq \gapDis \cdot \valMax, \quad \forall \ad,\genre,\slot. \]
    We therefore have $\gapVal \leq \gapDis$.
\end{proof}

\section{Experiment Details}\label{app:exp-setup}

\subsection{Human Experiment Instructions}\label{app:experiment-instructions}

\begin{figure}[htbp]
    \centering\includegraphics[width=1\linewidth]{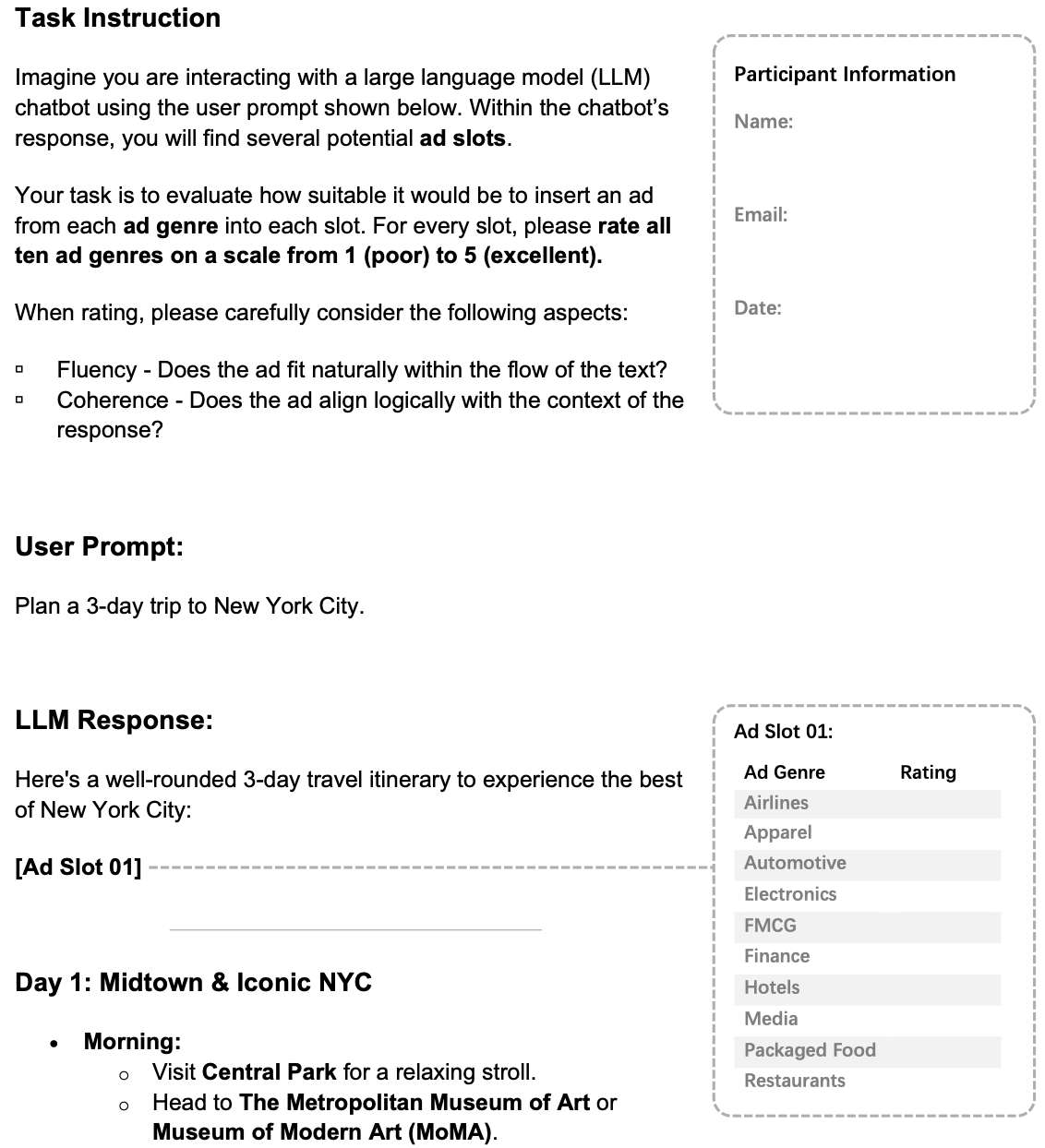}
    \caption{Example instruction for human participants.}
    \label{fig:human-instruction}
\end{figure}

\Cref{fig:human-instruction} shows an example of the instructions given to participants in the experiment described in \Cref{sec:experiment-coherence}. The full instructions are available in our anonymous repository at \url{https://anonymous.4open.science/r/Ad-Insertion-in-LLM-Generated-Responses-523F/}.

\subsection{Prompts for LLM-as-a-Judge}\label{app:prompts}

This section provides the prompt used for LLM-as-a-Judge coherence measurement in \Cref{sec:coherence,sec:experiment-coherence}.

------ System Prompt

\begin{lstlisting}
You are an expert in digital advertising and user experience. Your task is to rate how suitable different ad genres would be if inserted into a specific location within an LLM-generated response.
\end{lstlisting}

------ User Prompt

\begin{lstlisting}
Context: {}
User Query: {}

The LLM response contains an ad slot (marked as [Ad Slot]) where an advertisement could be inserted.

Here is the text surrounding Ad Slot:

{}
[Ad Slot] (THIS IS WHERE THE AD WOULD BE INSERTED)
{}

For each of the following ad genres, rate the suitability of inserting an ad from that genre into this slot on a scale from 1 (poor) to 5 (excellent).

When rating, consider:
- Fluency: Would the ad fit naturally within the flow of the text?
- Coherence: Would the ad align logically with the context of the response?

Here are the ad genres to rate:

Ad Genres:
1. Airlines - Examples: Flight deals, airline promotions
2. Apparel - Examples: Clothing, shoes, accessories
3. Automotive - Examples: Cars, motorcycles, electric vehicles
4. Electronics - Examples: Smartphones, computers, tablets
5. FMCG (Fast-Moving Consumer Goods) - Examples: Personal care products, household items
6. Finance - Examples: Banking services, insurance, credit cards
7. Hotels - Examples: Hotel chains, booking services
8. Media - Examples: Streaming services, social media platforms
9. Packaged Food - Examples: Snacks, beverages, prepared meals
10. Restaurants - Examples: Fast food, cafes, dining establishments

For each genre, provide:
1. A rating from 1-5
2. A brief explanation for your rating

Format your response as JSON with this structure:
{{
  "ratings": {{
    "Airlines": {{ "explanation": "...", "score": X }},
    "Apparel": {{ "explanation": "...", "score": X }},
    ...
  }}
}}

Ensure your ratings are justified based on the context and the natural flow of the text.
\end{lstlisting}

\end{document}